\documentclass[apj]{emulateapj}

\usepackage{apjfonts}
\usepackage{times}
\usepackage{amsmath}
\usepackage{graphicx}
\usepackage{natbib}
\slugcomment{Accepted for publication in ApJ}
\shorttitle{Recalibrating Single-Epoch Black Hole Mass Estimates}
\shortauthors{Park et al.}

\newcommand{\Hb}{H{$\beta$}}

\newcommand{\sigmaline}{$\sigma_{{\rm H}\beta}$}
\newcommand{\FWHM}{FWHM$_{{\rm H}\beta}$}

\newcommand{\mbh}{$M_{\rm BH}$}
\newcommand{\msigma}{$M_{\rm BH}-\sigma_{*}$}

\newcommand{\kms}{km~s$^{\rm -1}$}

\begin{document}
\title{The Lick AGN Monitoring Project: Recalibrating Single-Epoch Virial Black Hole Mass Estimates}
\author{Daeseong Park$^{1}$}
\author{Jong-Hak Woo$^{1}$\altaffilmark{,2}}
\author{Tommaso Treu$^{3}$\altaffilmark{,4}}
\author{Aaron J. Barth$^{5}$}
\author{Misty C. Bentz$^{6}$}
%in alphabetical order
\author{Vardha N. Bennert$^{3,7}$}
\author{Gabriela Canalizo$^{8}$}
\author{Alexei V. Filippenko$^{9}$}
\author{Elinor Gates$^{10}$}
\author{Jenny E. Greene$^{11}$}
\author{Matthew A. Malkan$^{12}$}
\author{Jonelle Walsh$^{5}$}
\affil{$^{1}$Astronomy Program, Department of Physics and Astronomy, Seoul National University, Seoul, 151-742, Republic of Korea\\
$^{3}$Department of Physics, University of California, Santa Barbara, CA 93106, USA\\
$^{5}$Department of Physics and Astronomy, 4129 Frederick Reines Hall, University of California, Irvine, CA 92697-4575, USA\\
$^{6}$Department of Physics and Astronomy, Georgia State University Atlanta, GA 30303, USA\\
$^{7}$Physics Department, California Polytechnic State University, San Luis Obispo, CA 93407, USA\\
$^{8}$Dept. of Physics and Astronomy, University of California, Riverside, 900 University Ave., Riverside, CA 92521, USA\\
$^{9}$Department of Astronomy, University of California, Berkeley, CA 94720-3411, USA\\
$^{10}$Lick Observatory, P.O. Box 85, Mount Hamilton, CA 95140, USA\\
$^{11}$Department of Astrophysical Sciences, Princeton University, Princeton, NJ 08544, USA.\\
$^{12}$Department of Physics and Astronomy, University of California, Los Angeles, CA 90024, USA\\}
\altaffiltext{2}{corresponding author; woo@astro.snu.ac.kr}
\altaffiltext{4}{Sloan Fellow; Packard Fellow.}

%%%%%%%%%%%%%%%%%%%%%%%%%%%%%%%%%%%%%%%%%%%%%%%%%%%%%%%%%%%%%%%%%%%%%%%%%%%
\begin{abstract}
We investigate the calibration and uncertainties of black hole mass
estimates based on the single-epoch (SE) method, using homogeneous and
high-quality multi-epoch spectra obtained by the Lick Active Galactic
Nucleus (AGN) Monitoring Project for 9 local Seyfert~1 galaxies with
black hole masses $<10^{8}~{\rm M}_{\odot}$.  By decomposing the
spectra into their AGN and stellar components, we study the
variability of the single-epoch H$\beta$ line width (full width at
half-maximum intensity, \FWHM; or dispersion, \sigmaline) and of the
AGN continuum luminosity at 5100~\AA~($L_{\rm 5100}$).  From the
distribution of the ``virial products'' ($\propto$~\FWHM$^2$ $L_{\rm
5100}^{0.5}$ or \sigmaline$^2$ $L_{\rm 5100}^{0.5}$) measured from SE
spectra, we estimate the uncertainty due to the combined variability
as $\sim0.05$ dex (12\%).  This is subdominant with respect to the
total uncertainty in SE mass estimates, which is dominated by
uncertainties in the size-luminosity relation and virial coefficient,
and is estimated to be $\sim 0.46$ dex (factor of $\sim3$).  By
comparing the \Hb~line profile of the SE, mean, and root-mean-square
(rms) spectra, we find that the \Hb\ line is broader in the mean (and
SE) spectra than in the rms spectra by $\sim0.1$ dex (25\%) for our
sample with \FWHM\ $<3000$ \kms. This result is at variance with
larger mass black holes where the difference is typically found to be
much less than $0.1$ dex.  To correct for this systematic difference
of the \Hb\ line profile, we introduce a line-width dependent virial
factor, resulting in a recalibration of SE black hole mass estimators
for low-mass AGNs.
\end{abstract}

%%%%%%%%%%%%%%%%%%%%%%%%%%%%%%%%%%%%%%%%%%%%%%%%%%%%%%%%%%%%%%%%%%%%%%%%%%%
% Introduction
%%%%%%%%%%%%%%%%%%%%%%%%%%%%%%%%%%%%%%%%%%%%%%%%%%%%%%%%%%%%%%%%%%%%%%%%%%%
\section{INTRODUCTION}  
\label{section:intro}
Supermassive black holes (BHs) are believed to play a key role in
galaxy evolution. Evidence for this connection comes from the tight
correlations observed in the local universe between BH masses and the
global properties of their host galaxies (Magorrian et al. 1998;
Ferrarese \& Merritt 2000; Gebhardt et al. 2000; G\"ultekin et
al. 2009; Bentz et al. 2009a; Woo et al. 2010). Establishing the cosmic
evolution of these correlations is a powerful way to understand the
feedback mechanisms connecting BHs and galaxies (e.g., Kauffmann \&
Haehnelt 2000; Robertson et al. 2006; Hopkins et al. 2006).  Recent
observational studies have found that these correlations may evolve
over cosmic time, in the sense that BHs of a given mass appeared to
live in smaller galaxies in the past (e.g., Woo et al. 2006; Peng et
al. 2006; Treu et al. 2007; Woo et al.\ 2008; Merloni et al.  2010;
Decarli et al.\ 2010; Bennert et al.\ 2010).  

In order to investigate the nature of BH-galaxy coevolution, as well
as virtually all aspects of active galactic nucleus (AGN) physics
(e.g., Woo \& Urry 2002; Kollmeier et al. 2006; Davis et al. 2007), BH
masses must be accurately determined at large distances.  Dynamical
methods based on high angular resolution kinematics of stars and gas
are the most common approach to measuring masses of quiescent BHs
(e.g., Kormendy \& Gebhardt 2001; Ferrarese \& Ford 2005). However,
owing to the parsec-size scale of the sphere of the influence of
typical BHs, they are limited to galaxies within a distance of few
tens of Mpc with current technology.

In the case of BHs powering an AGN, the presence of a variable
broad-line region (BLR) provides an alternative way that is in
principle applicable to much larger distances. The geometry and
kinematics of the BLR gas can be mapped in the time domain using the
so-called reverberation (or echo) mapping technique (Blandford \&
McKee 1982; Peterson 1993). In turn, these quantities can be converted
into BH mass estimates under appropriate assumptions about the
dynamics of the system (Peterson 1993; Pancoast, Brewer, \& Treu
2011). Estimators of the form \mbh\ $\propto R_{\rm BLR}~ V^2$, where
$R_{\rm BLR}$ and $V$ are (respectively) size and velocity estimators
of the BLR, are often referred to as ``virial'' mass estimators.
However, due to the observational challenges of reverberation mapping
campaigns, fewer than 50 BH masses have been measured to date using
this technique (Wandel, Peterson, \& Malkan 1999; Kaspi et al. 2000;
Peterson et al. 2004; Bentz et al. 2009c; Denney et al 2009; Barth et
al. 2011).  

In light of the scientific importance of determining BH masses, it is
critical to develop alternative BH mass estimators that are
observationally less demanding. A popular BH mass estimator, based on
the results of reverberation mapping studies, is the so-called
single-epoch (SE) method. It exploits the empirical correlation
between the size of the BLR and the AGN continuum luminosity ($R_{\rm
BLR} \propto L^{\alpha}$, with $\alpha
\approx 0.5$), as expected from the photoionization model predictions
(Wandel, Peterson, \& Malkan 1999; Kaspi et al. 2000, 2005; Bentz et
al. 2006, 2009b), to bypass the expense of a monitoring
campaign. Thus, the AGN luminosity is used as a proxy for the BLR size
and, in combination with the square of a velocity estimate from a
broad line, to estimate BH masses from single spectroscopic
observations.  Typically, SE mass estimators are based on
optical/ultraviolet lines (e.g., H$\beta$ or \ion{Mg}{2}) and
optical/ultraviolet continuum luminosity (e.g., at 5100~\AA\ or
3000~\AA). A summary and cross-calibration of commonly adopted recipes
is given by McGill et al. (2008).  

Due to its convenience, the SE method has been widely applied from the
study of BH demographics (e.g., Shen et al. 2008; Fine et al. 2008) to
the characterization of galaxy-AGN scaling relations at low and high
redshift (e.g., Treu et al. 2004; Barth et al. 2005; Greene \& Ho
2006; Woo et al. 2006; Bennert et al.\ 2010, Bennert et al.\ 2011a).
For this reason it is of paramount importance to quantify, understand,
and (possibly) correct for random and systematic uncertainties in the
method.  In addition to the random and systematic errors, selection
bias can play a role in studying statistical properties of AGN samples
selected from a flux-limited survey since BH mass from SE data is
proportional to the AGN continuum luminosity at 5100\AA~($L_{5100}$) 
(e.g., Lauer et al. 2007; Treu et al. 2007; Shen \& Kelly 2010). 
Naturally, the strength of the selection bias
depends on the uncertainty of the SE mass estimates, providing another
compelling reason to quantify it accurately.  

The largest uncertainty comes from the unknown ``virial'' factor $f$,
connecting the observable size and velocity to the actual BH mass,
\mbh\ $ \equiv f R_{\rm BLR}~ V^2 /G$, where $G$ is the gravitational
constant. In general, $f$ cannot be determined for individual sources
due to limited spatial information except a few cases
(Davies et al. 2006, Onken et al. 2007; Hicks \& Malkan 2008;
see, however, Brewer et al. 2011 and references therein).  
Therefore, an average virial factor is
typically applied. This average is determined by forcing active and
quiescent galaxies to obey the same BH mass-galaxy velocity
dispersion (\msigma) relation (Onken et al. 2004; Woo et al. 2010),
even though the virial factor of individual AGNs may be different from
the mean value.  Thus, using an average virial factor introduces an
uncertainty in the SE mass estimates.  It is not known precisely how
large the uncertainty of the virial factor is (see Collin et
al. 2006), and whether this uncertainty is stochastic (random) or has
a systematic component that can be reduced using additional
observables.  An upper limit to the uncertainty is derived from the
intrinsic scatter of the AGN \msigma\ relation (0.43 dex; Woo et
al. 2010), assuming that the samples used to calibrate $f$ are
representative of the class of broad-line AGNs targeted for the SE
study.

A second source of uncertainty is the variability of AGNs: line width
and continuum luminosity will vary as a function of time, while the BH
mass is not expected to change significantly over time scales of order
a few years.  Thus, AGN variability introduces an uncertainty in the
SE mass estimates, which is believed to be stochastic in
nature. Previous studies based on multi-epoch spectra reported that
the random error due to the variability is $\sim$15--25\% (e.g., Woo
et al. 2007; Denney et al. 2009).

A third source of error is the intrinsic scatter in the
size-luminosity relation used to infer the size of the BLR.  Recent
studies, based on reverberation mapping results and {\it Hubble Space
Telescope (HST)} imaging analysis, report $\sim$40\% scatter
in the size-luminosity relation (Bentz et al. 2009b).

A fourth source of uncertainty in SE mass estimates is due to
differences in the BLR line profile as measured from SE
spectra and those measured from root-mean-square (rms) spectra.
In reverberation mapping studies, BH mass determinations 
rely on the line width measured from the rms spectra, 
which reflect the varying part of the line profile.
In contrast, for SE mass determinations line widths are measured from single spectra
since such equivalent measurements as in the rms spectra are not available.
Thus, it is necessary to investigate and quantify the line-width difference 
between SE and rms spectra.
Previous studies based on multi-epoch data showed that the H$\beta$ 
line widths in the mean spectra are broader than those in the rms spectra
(e.g., Sergeev et al. 1999; Shapovalova et al. 2004; Collin et al. 2006; Denney et al. 2009).
The difference is presumably due to the different kinematics of the gas responding 
over various time scales, indicating that a different normalization is required
in order to consistently estimate virial masses based on the SE method.

In this work, we focus on the uncertainties of SE mass estimators due
to the variability, and those due to differences in line profiles.  By
comparing measurements from single-epoch, mean, and rms spectra using
the high-quality multi-epoch spectra of 9 local Seyfert galaxies in
the relatively unexplored regime of low-mass BHs from the Lick AGN
monitoring project (\citealt{bentz09c}), we provide new quantitative
estimates for the uncertainties and recipes to correct for them.  The
paper is organized as follows. In \S \ref{section:obs}, we describe
the observations and data reduction. Section~\ref{section:meas}
discusses the measurement method for SE spectra as well as for mean
and rms spectra.  In \S~\ref{section:analresul}, we present the main
results including a test of the virial assumption, a quantification of
the random errors due to AGN variability, and the systematic
differences in line width between SE and rms spectra. We also present
a recalibration of standard recipes that corrects for the systematic
differences.  We conclude and summarize our findings in
\S~\ref{section:diss}.  Throughout the paper we adopt the following
cosmological parameters to calculate distances: $H_0 =
70$~km~s$^{-1}$~Mpc$^{-1}$, $\Omega_{\rm m} = 0.30$, and
$\Omega_{\Lambda} = 0.70$.
%%%%%%%%%%%%%%%%%%%%%%%%%%%%%%%%%%%%%%%%%%%%%%%%%%%%%%%%%%%%%%%%%%%%%%%%%%%
% Observations and Data Reduction
%%%%%%%%%%%%%%%%%%%%%%%%%%%%%%%%%%%%%%%%%%%%%%%%%%%%%%%%%%%%%%%%%%%%%%%%%%%
\section{OBSERVATIONS AND DATA REDUCTION} \label{section:obs}

We use the homogeneous and high-quality multi-epoch spectra from the
Lick AGN monitoring project (LAMP; \citealt{bentz09c}), which was
designed to measure the reverberation time scales of 13 local Seyfert
1 galaxies.  Here, we briefly summarize the observations and data
reduction.

The LAMP campaign was carried out using the Kast spectrograph at the
3-m Shane telescope at the Lick Observatory in Spring 2008.  Among 13
Seyfert 1 galaxies, we selected 9 objects for which the \Hb\ line
variability was sufficiently large to measure the reverberation time lag
(Bentz et al. 2009c). During the LAMP campaign, each object was
observed multiple times (43 to 51 epochs with an average of 47),
enabling us to construct high-quality mean and r.m.s. spectra.

After performing standard spectroscopic reductions using
IRAF\footnote{IRAF is distributed by the National Optical Astronomy
Observatories, which are operated by the Association of Universities
for Research in Astronomy, Inc., under cooperative agreement with the
National Science Foundation (NSF).}, one-dimensional spectra were
extracted with an aperture window of 13 pixels (10\farcs1).  Flux
calibrations utilized nightly spectra of
spectrophotometric standard stars.  As described by Bentz et
al. (2009c), the spectral rescaling was performed using the algorithm
of \citet{vangroningen92} in order to mitigate the effects of slit
loss, variable seeing, and transparency.  By rescaling, shifting, and
smoothing each spectrum, the algorithm minimizes the difference of
flux of the [\ion{O}{3}] lines between each spectrum and a reference
spectrum created from the mean of individual spectra.  The quality of
each individual spectrum is sufficient to perform SE measurements
(average signal-to-noise ratio S/N $\approx 66$ per pixel at rest-frame 5100~\AA).
%%%%%%%%%%%%%%%%%%%%%%%%%%%%%%%%%%%%%%%%%%%%%%%%%%%%%%%%%%%%%%%%%%%%%%%%%%%
% Measurements
%%%%%%%%%%%%%%%%%%%%%%%%%%%%%%%%%%%%%%%%%%%%%%%%%%%%%%%%%%%%%%%%%%%%%%%%%%%
\section{MEASUREMENTS} \label{section:meas}

Two quantities, the line width and the continuum luminosity, are required
to determine \mbh\ using single spectra. Uniform and consistent
analysis is crucial for investigating systematic uncertainties and
minimizing additional errors. In this section, we present the
multi-component spectral fitting process and describe the measurements
using single-epoch, mean, and rms spectra.

\begin{figure*}%[h]
\centering
    \includegraphics[width=\textwidth]{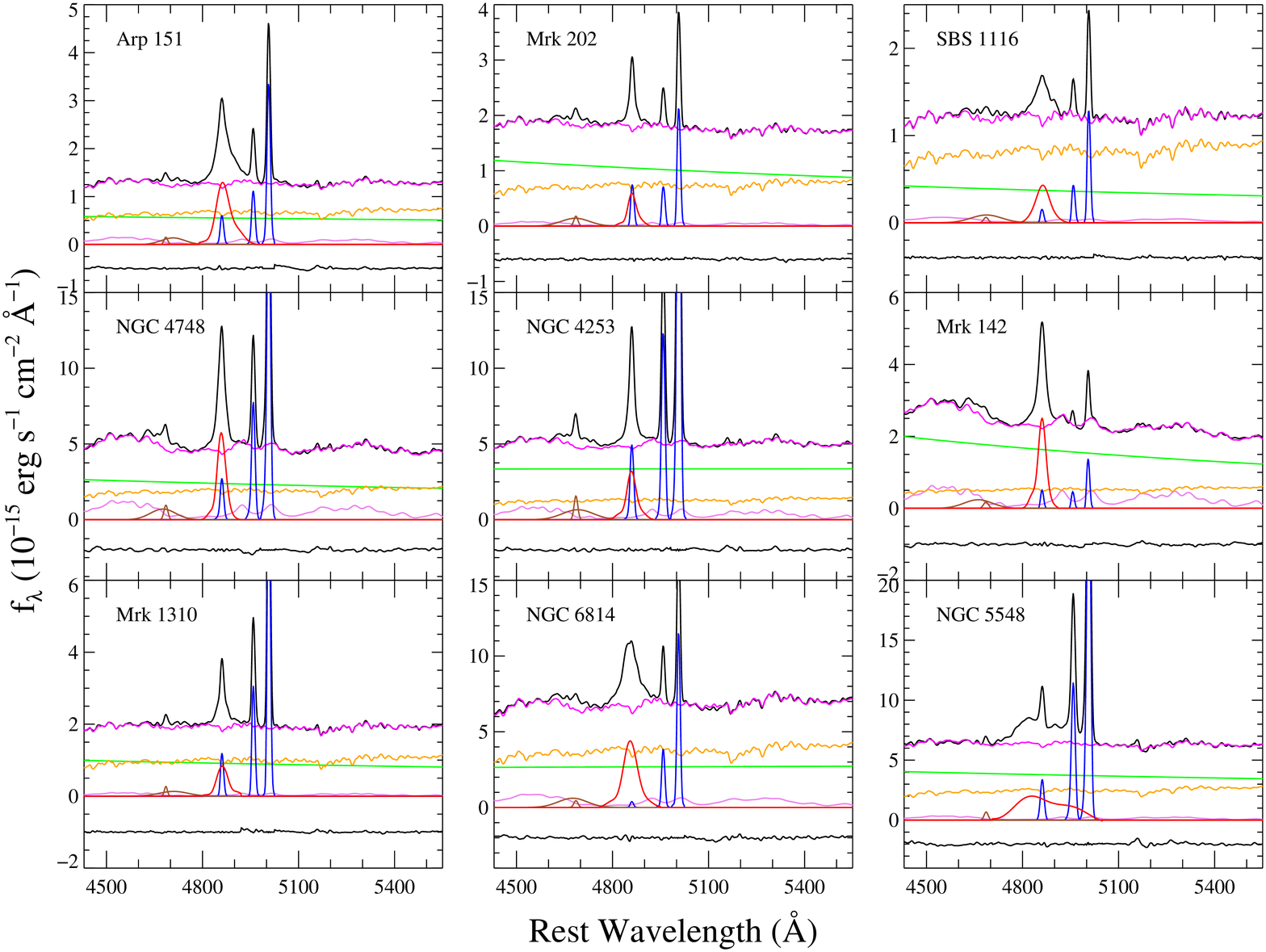}
    \caption{Multi-component spectral fitting in the mean spectra.
    The mean spectra of all 9 Seyfert galaxies are presented along
    with multi-component models. In each panel, observed spectra
    (black) and the continuum$+$\ion{Fe}{2}$+$stellar best-fit model
    (magenta) are shown in the upper part, and the best-fit power-law
    continuum (green), stellar spectrum (yellow), and \ion{Fe}{2}
    template (violet) models are presented in the middle part.  Three
    narrow lines [\Hb, [\ion{O}{3}] $\lambda\lambda 4959,5007$
    (blue)], broad \Hb\ (red), and the broad and narrow \ion{He}{2}
    $\lambda4686$ components (brown) are presented in the bottom part.
    The residuals (black), representing the difference between the
    observed spectra and the sum of all model components, are
    arbitrarily shifted downward for clarity.
    \label{fig:all_fit_mean}}
\end{figure*}

\subsection{Multi-Component Fitting}

To measure the line width of H$\beta$ and the continuum luminosity at
5100~\AA, we follow the procedure given by \citet{woo06} and
\citet{mcgill08}, but with significant modifications as described below 
(cf., McLure \& Dunlop 2004; Dietrich et al. 2005; Denney et
al. 2009).  The multi-component fitting processes were carried out in
a simultaneous and automated fashion, using the nonlinear
Levenberg-Marquardt least-squares fitting routine \texttt{mpfit}
\citep{Markwardt09} in IDL.

First, all single spectra were converted to the rest frame.  Second,
we modelled the observed continuum with three components: the
featureless AGN continuum, the \ion{Fe}{2} emission blends, and the
host-galaxy starlight, using respectively a single power-law
continuum, an \ion{Fe}{2} template from \citet{Boroson92}, and a
host-galaxy template from \citet{Bruzual03}.  A simple stellar
population synthesis model with solar metallicity and age of 11~Gyr
was found to reproduce the observed stellar lines reasonably well (see
Figure 1). The \ion{Fe}{2} emission blends and the host-galaxy
template were convolved with appropriate Gaussian velocities to
reproduce kinetic and instrumental broadening during the fitting
process as described below.  The best continuum models were determined
based on the $\chi^2$ statistic in the regions 4430--4600~\AA\ and
5080--5550~\AA\ where \ion{Fe}{2} emission dominates.  The three
components were varied simultaneously with six free parameters: the
normalization and the slope of the power-law continuum, the strength
and the broadening velocity of the \ion{Fe}{2}, and the line strength
and the velocity dispersion of the host-galaxy templates.  We masked
out the typical weak AGN narrow emission lines (e.g.,
\ion{He}{1} $\lambda4471$, [\ion{Fe}{7}] $\lambda5160$, [\ion{N}{1}]
$\lambda5201$ , [\ion{Ca}{5}] $\lambda5310$; \citealt{vanden01})
during the fitting process. The best-fit continuum models (the power-law component 
+ the \ion{Fe}{2} template + the host-galaxy template) were
subtracted from each spectrum, leaving the broad and narrow AGN
emission lines.

\begin{deluxetable}{lcc}
\tablecolumns{3}
\tablewidth{0pt}
\tabletypesize{\footnotesize}
\tablecaption{H$\beta$ Integration Ranges, and H$\beta$ narrow ratios}
\tablehead{
\colhead{Object} &
\colhead{H$\beta_{\rm BC}$ Line Ranges} &
\colhead{$f$(H$\beta_{\rm NC}$)/$f$([\ion{O}{3}] $\lambda 5007$)} \\
\colhead{} &
\colhead{(\AA)} &
\colhead{}}
\startdata
Arp\,151        & 4790--4980 & 0.18   \\
NGC\,4748       & 4790--4920 & 0.12   \\
Mrk\,1310       & 4800--4920 & 0.13   \\
Mrk\,202        & 4810--4920 & 0.35   \\
NGC\,4253 (Mrk\,766)       & 4790--4930 & 0.13   \\
NGC\,6814       & 4760--4950 & 0.03   \\
SBS\,1116+583A  & 4795--4940 & 0.12   \\
Mrk\,142        & 4790--4910 & 0.37   \\
NGC\,5548       & 4705--5040 & 0.10   %\\
\enddata
\tablecomments{All values in the table are given in the rest frame.}
\label{table:range}
\end{deluxetable}

Third, we subtracted the narrow lines around the H$\beta$ region
before fitting the broad component. We first made a template for the
H$\beta$ narrow-line profile by fitting a tenth-order Gauss-Hermite
series (cf., van der Marel \& Franx 1993) model to the [\ion{O}{3}]
$\lambda5007$ line. We then subtracted the [\ion{O}{3}] $\lambda4959$
line by blueshifting and scaling the flux of the template by
$1/3$. The H$\beta$ narrow line was also subtracted by scaling the
[\ion{O}{3}] $\lambda5007$ line.  The ratios of the narrow H$\beta$ to
[\ion{O}{3}] $\lambda5007$ were determined from the $\chi^2$
minimization in the mean spectra and then forced to be the same for
all SE spectra of each object.  Applied scaling ratios for the
H$\beta$ narrow component range from 0.03 to 0.37
(Table~\ref{table:range}).

Lastly, we modelled the broad component of the H$\beta$ line using a
sixth-order Gauss-Hermite series. We also used a two-component
Gaussian model to describe the broad and narrow components of the
\ion{He}{2} $\lambda4686$ emission line whenever it affected the blue wing
of the H$\beta$ profile.  Figure~\ref{fig:all_fit_mean} shows the
fitting results for the mean spectra.

\begin{figure}     
\centering
    \includegraphics[width=0.475\textwidth]{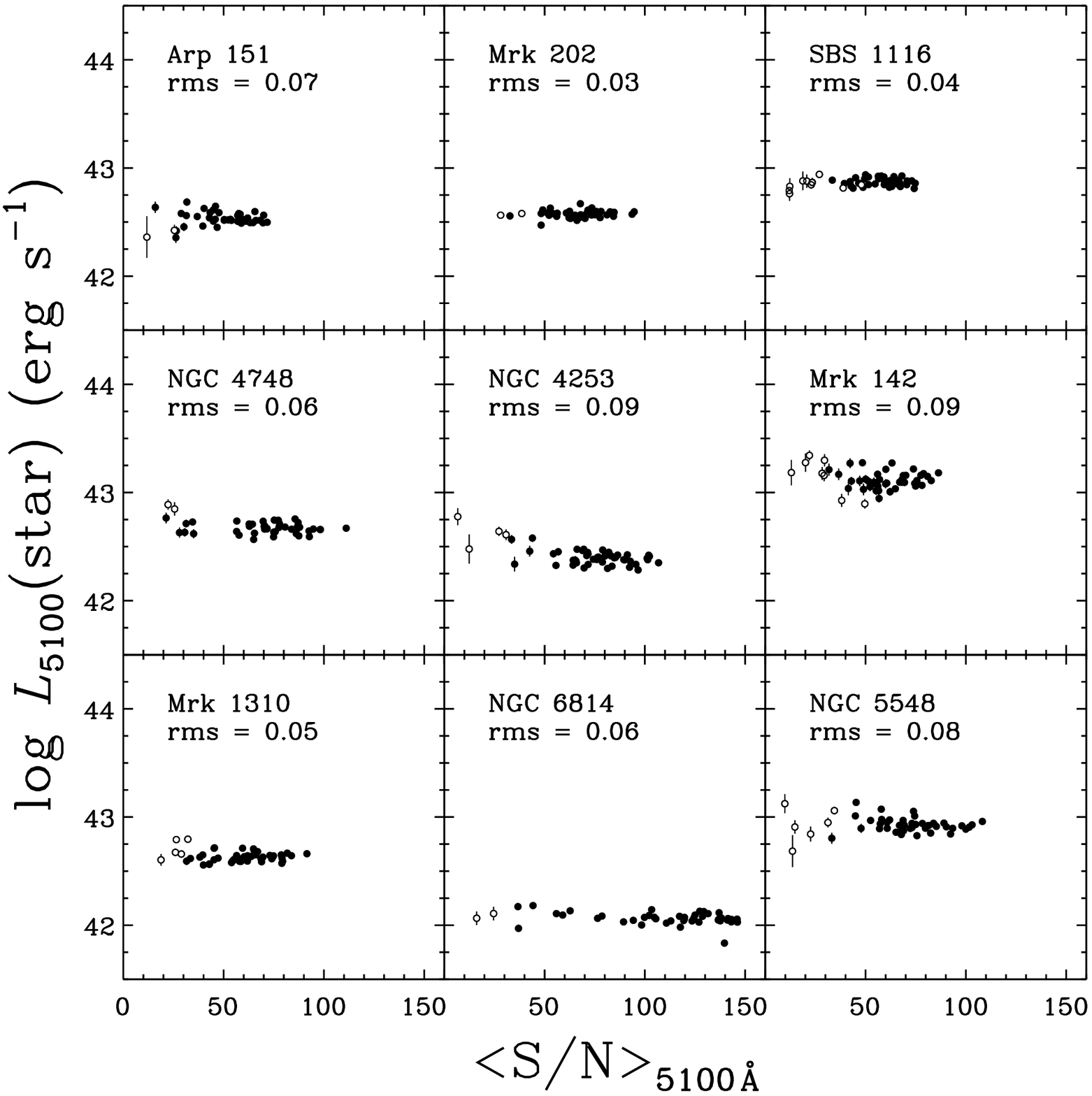}
    \caption{The host-galaxy luminosity at 5100~\AA, measured from each
    SE spectrum based on the spectral decomposition analysis,
    as a function of S/N.  A few low-S/N spectra (open circles) are
    removed from further analysis to avoid biases due to low-quality
    data. Measurement uncertainties estimated from the method given in
    \S 3.2.3 are expressed by vertical error bars.
    \label{fig:HGfraction}}
\end{figure}

\subsection{Single-Epoch Spectra}

We performed the multi-component fitting procedure using individual SE
spectra, and measured the line width and continuum luminosity for each
epoch.  The vast majority of SE spectra have sufficiently high quality
to perform the analysis (S/N $\approx 66$ per pixel at rest-frame
5100~\AA). However, a small fraction of spectra have significantly
lower S/N owing to bad weather during the LAMP monitoring
campaign. In addition, there are a few epochs with artificial signatures,
such as bad pixels, abnormal curvature, or fluctuations in the reduced spectra.
Those SE spectra were discarded to avoid possible biases due
to much larger measurement errors (see Fig. \ref{fig:HGfraction}).  
On average, four bad epochs out of 47 nights were removed for each
object, except for SBS~1116, for which 11 epochs were eliminated
because of a defect between the H$\beta$ and [\ion{O}{3}]
$\lambda4959$ lines due to bad pixels in the detector.

\begin{figure}
\centering
    \includegraphics[width=0.45\textwidth]{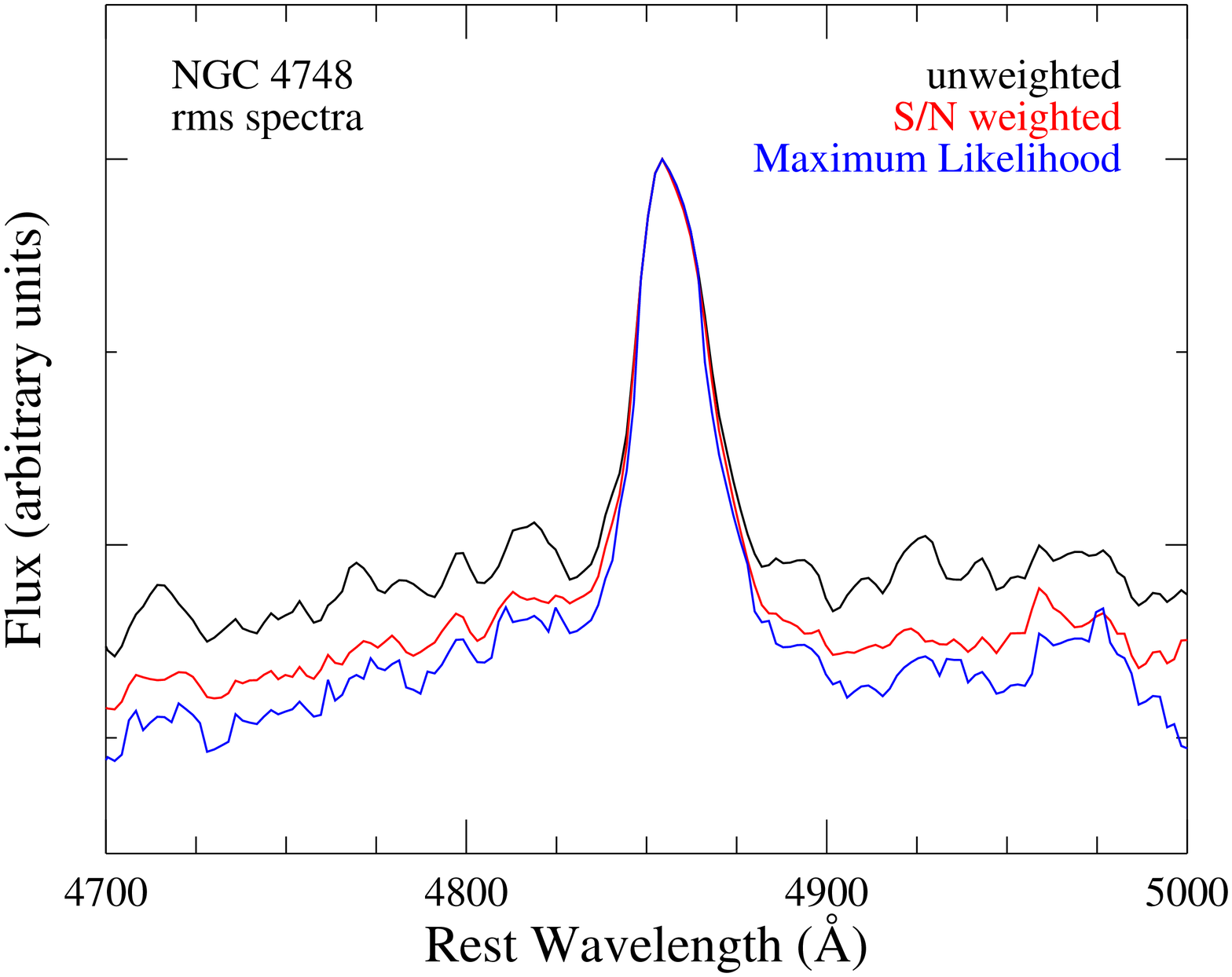}
    \caption{Comparison of the rms spectra of NGC\,4748 generated
    with three different methods: unweighted rms (black), S/N weighted
    (red), and maximum likelihood (blue).  
    For this object, two bad epochs with low-S/N data were removed as
    described in \S~3.2.
    \label{fig:compare_rms}}
\end{figure}

\subsubsection{Emission-Line Width}

We measured the full width at half-maximum intensity (\FWHM) and the
dispersion (\sigmaline) \citep[the second moment of line
profile;][]{peterson04} of the broad component of \Hb\ directly from
the data as well as from the fits to the continuum-subtracted
spectra. Line-width measurements are corrected for the instrumental
resolution in a standard way (Barth et al. 2002; Woo et al. 2004;
Bentz et al. 2009c), by subtracting in quadrature the instrumental
resolution (Table 11 of Bentz et al. 2009c) from the measured line width.

By comparing line widths measured from Gauss-Hermite series fits 
with those directly measured from the data, we found less than a 3\% 
systematic difference (with considerable rms scatter of $\sim 5$\%)
as expected given the high S/N of individual spectra. 
The small systematic trend between \FWHM~and \sigmaline~shows opposite directions.  
In the case of \FWHM, the measurements from the fit were $2.6\pm0.2$\% larger 
than those from the data while \sigmaline~measurements from the fit
were $1.9\pm0.1$\% smaller than those from the data, showing a trend consistent
with that reported by Denney et al. (2009).
For consistency with other studies on the reverberation and single-epoch masses, 
we focus on the line-width measurements 
from the fits in the rest of the paper unless explicitly noted.
%since this method in general provides more reliable results in lower
%S/N spectra (down to $\sim 10$; see Denney et al. 2009).

%%%%%%%%%%%%%%%%%%%%%%%%%%%%%%%%%%%%%%%%%%%%%%%%%%%%%%%%%%%%%%%%%%
%\setlength{\tabcolsep}{0.005in}
\begin{deluxetable}{lcccc}
%\rotate {}
\tablewidth{0pt}
\tablecolumns{5}
\tabletypesize{\footnotesize}
%\tablecaption{Continuum luminosities and host-galaxy contributions from spectral decomposition in the mean spectra}
\tablecaption{Mean continuum luminosities and host-galaxy contributions}
\tablehead{
\colhead{Object}    &
\colhead{$\lambda L_{\lambda}(\mathrm{tot})$}     &
\colhead{$\lambda L_{\lambda}(\mathrm{AGN})$}    &
\colhead{$\lambda L_{\lambda}(\mathrm{star})$}    &
%\colhead{$L_{\mathrm{star}}$/$L_{\mathrm{tot}}$}
%\colhead{$\lambda L_{\lambda}(\mathrm{star})$/$\lambda L_{\lambda}(\mathrm{tot})$}
\colhead{\scalebox{1.2}{$\frac{{\lambda L_{\lambda}(\mathrm{star})}}{{\lambda L_{\lambda}(\mathrm{tot})}}$}}
%\multirow{2}{*}{$\frac{{\lambda L_{\lambda}(\mathrm{star})}}{{\lambda L_{\lambda}(\mathrm{tot})}}$}
\\
\colhead{} &
\colhead{(10$^{42}$ erg s$^{-1}$)} &
\colhead{(10$^{42}$ erg s$^{-1}$)} &
\colhead{(10$^{42}$ erg s$^{-1}$)} &
\colhead{}
\\
\colhead{(1)} &
\colhead{(2)} &
\colhead{(3)} &
\colhead{(4)} &
\colhead{(5)}
}

\startdata
Arp\,151    & $ 6.11$ & $ 2.76$ & $ 3.42$ & $0.56$   \\
NGC\,4748   & $10.05$ & $ 5.55$ & $ 4.69$ & $0.47$   \\
Mrk\,1310   & $ 8.10$ & $ 3.79$ & $ 4.32$ & $0.53$   \\
Mrk\,202    & $ 8.78$ & $ 5.02$ & $ 3.75$ & $0.43$   \\
NGC\,4253   & $ 8.87$ & $ 6.41$ & $ 2.47$ & $0.28$   \\
NGC\,6814   & $ 1.93$ & $ 0.83$ & $ 1.15$ & $0.59$   \\
SBS\,1116+583A  & $10.52$ & $ 3.15$ & $ 7.44$ & $0.71$   \\
Mrk\,142    & $47.89$ & $35.64$ & $12.73$ & $0.27$   \\
NGC\,5548   & $20.86$ & $12.42$ & $ 8.37$ & $0.40$   %\\
\enddata
\label{tab:luminosity_from_fit}
\tablecomments{
Col. (1): object name.
Col. (2): the total continuum luminosity at 5100~\AA.
Col. (3): the AGN luminosity estimated from the power-law continuum fit.
Col. (4): the host-galaxy luminosity estimated from the host-galaxy template fit.
Col. (5): the host-galaxy fraction.
}
\end{deluxetable}

\subsubsection{Continuum Luminosity} \label{section:meas:conti}

We measured the monochromatic continuum luminosity at 5100~\AA\
from the observed spectra at each epoch by calculating the average
flux in the rest-frame 5080--5120~\AA\ region.  The luminosity at
5100~\AA\ (total luminosity, $L_{\rm 5100, t}$) is strongly
contaminated by the host-galaxy starlight when the AGN luminosity is
comparable to or smaller than the host-galaxy stellar luminosity as in the
Seyfert galaxies in our sample.

To obtain the AGN continuum luminosity (nuclear luminosity, $L_{\rm 5100,
n}$), the host-galaxy contribution to the total luminosity should be
subtracted from the measured total luminosity.  In principle, the 
host-galaxy luminosity can be determined by separating a stellar component
from a point source using surface brightness fitting analysis based on
a high-resolution image. Such an analysis is in progress based on the
{\it HST} WFC3 images of the LAMP sample (GO-11662, PI. Bentz). For this
paper, however, we used the information obtained from the spectral
decomposition. We note that, although the host-galaxy flux should be
constant, the amount of host-galaxy contribution to the total flux can
vary in each epoch's spectrum because of seeing variations and
miscentering in the slit.  Thus, the nuclear luminosity, $L_{\rm
5100, n}$, needs to be estimated for each individual spectrum from which
\ion{Fe}{2} and starlight have been subtracted.

Figure~\ref{fig:HGfraction} shows the starlight luminosity measured
from each SE spectrum as a function of S/N.  As expected, the
starlight is not constant due to the effects of seeing and
miscentering.  The variability ranges from 10\% to 20\% with an
average of $0.06\pm0.01$ dex. These results underscore the importance
of subtracting the host-galaxy starlight in making the rms
spectra. Otherwise, the rms spectra may contain a contribution from
the variable amount of starlight observed through the slit (see \S~3.3
and Figure~\ref{fig:rms_stellar}).

As a consistency check, we directly compare the host-galaxy flux of NGC\,5548 
measured from our spectral decomposition with that from the {\it HST} imaging analysis 
as similarly done by Bentz et al. (2009b).
In order to calculate the amount of light observed through the spectroscopic
aperture, we used an aperture size of $4'' \times 10\farcs1$ 
as used in the LAMP spectroscopy analysis, 
after smearing the point-spread-function (PSF) subtracted {\it HST} image with a $2''$ Gaussian seeing disk.
The host-galaxy flux of NGC\,5548 based on the spectral decomposition
is $2.47\times 10^{-15}$ erg~s$^{-1}$~cm$^{-2}$~\AA$^{-1}$,
while the {\it HST} imaging-based galaxy flux is 
$2.73\times 10^{-15}$ erg s$^{-1}$~cm$^{-2}$~\AA$^{-1}$.
Thus, the small difference ($\sim 10$\%) between the two analyses 
shows the consistency in host-galaxy flux measurements. 
When we use a smaller seeing disk (e.g., a 1\farcs5 Gaussian disk),
the host-galaxy flux measured from the {\it HST} imaging analysis increases by $\sim 13$\%,
indicating that the actual seeing size will slightly change the host-galaxy 
flux measurements.

\subsubsection{Error Estimation}

To estimate the uncertainties of the line-width and luminosity
measurements from SE spectra, we adopted the Monte Carlo flux
randomization method \citep[e.g.,][]{bentz09c,shen11}. First, we
generated 50 mock spectra for each observed spectrum by adding
Gaussian random noise based on the flux errors at each spectral pixel.
Then we measured the line widths and AGN luminosities from the
simulated spectra using the method described in \S 3.2.1 and \S 3.2.2.
We adopted the standard deviation of the distribution of measurements
from 50 mock spectra as the measurement uncertainty.  For a
consistency check, we increased the number of mock spectra up to 100
and found that the results remain the same.  In the case of the total
luminosity ($L_{\rm 5100,t}$), we measured the uncertainty as the
square root of the quadratic sum of the standard deviation of fluxes
and average flux errors in the continuum-flux window.

\subsection{Mean and RMS Spectra}

In this section, we describe the process of generating mean and rms
spectra, and present the method for measuring the line width and
continuum luminosity.  The mean spectra are representative of all
single spectra, thus they are useful to constrain the random errors of
measurements from single-epoch spectra. In contrast, reverberation
mapping studies generally use rms spectra to map the geometry and
kinematics of the same gas that responds to the continuum variation.
By comparing the line profiles between rms and single spectra, one can
investigate any systematic differences of the corresponding line
widths, and therefore improve the calibration of BH mass estimators.

\subsubsection{Method}

We generated mean and rms spectra for each object using the following
equations:
\begin{equation}
\left\langle {f(\lambda )} \right\rangle = \frac{1}{N}\sum\limits_{i
= 1}^N {f_i (\lambda )},
\end{equation}
\begin{equation}
{\rm rms}(\lambda ) = \sqrt {\frac{1}{{N - 1}}\sum\limits_{i = 1}^N
{\left[ {f_i (\lambda ) - \left\langle {f(\lambda )} \right\rangle}
\right]^2 } },
\end{equation}
where $f_i (\lambda )$ is the flux of $i$-th SE spectrum
(out of $N$ spectra). 

The unweighted rms spectra can be biased by low-S/N spectra, often
showing peaky residual features in the continuum.  These spurious
features in continuum can affect the wings of the emission lines and
therefore the measurement of line dispersion.  To mitigate this effect
it is best to consider more robust procedures. We considered the
following two schemes.  First, we used the S/N as a weight, with the
following equations:
\begin{equation}
\left\langle {f^{w}(\lambda )} \right\rangle = \sum\limits_{i = 1}^N
{w_i f_i (\lambda )},
\end{equation}
\begin{equation}
{\rm rms}^{w} (\lambda ) = \sqrt {\frac{1}
{{1 - \sum\limits_{i = 1}^N {w_i^2 } }}\sum\limits_{i = 1}^N {w_i \left[ {f_i (\lambda ) -
\left\langle {f^{w}(\lambda )} \right\rangle} \right]^2 } },
\end{equation}
where $w_i$ is the normalized S/N weight defined by
\begin{equation}
w_i  = \frac{{\left( {S_i /N_i } \right)}}{{\sum\limits_{i = 1}^N
{\left( {S_i /N_i } \right)} }}.
\end{equation}

Alternatively, we considered the maximum likelihood method. Assuming
Gaussian errors, the logarithm of the likelihood function (up to a
normalization constant) is given by
%\begin{equation}
%2\ln \mathcal{L} =  - \sum\limits_{i = 1}^N {\ln \varepsilon
%_{tot,i}^2  - } \sum\limits_{i = 1}^N {\frac{{\left( {f_i^{mean}  -
%f_i } \right)^2 }}{{\varepsilon _{tot,i}^2 }}},
%\end{equation}
%\begin{equation}
%2\ln \mathcal{L} =  - \sum\limits_{i = 1}^N {\left\{ {\ln \epsilon
%_{tot,i}^2(\lambda)  + \frac{{\left[ {f_i (\lambda ) - {\rm
%mean}(\lambda )} \right]^2 }}{{\epsilon _{tot,i}^2(\lambda) }}}
%\right\}}
%\end{equation}
%\begin{equation}
%\varepsilon ^2  = \varepsilon _{flux}^2  + \sigma _{rms}^2
%\end{equation}
%
\begin{equation}
2\ln \mathcal{L} =  - \sum\limits_{i = 1}^N {\ln \epsilon _{{\rm
tot}, i}^2 (\lambda )} - \sum\limits_{i = 1}^N {\frac{{\left[ { f_i
(\lambda ) - \left\langle {f(\lambda )} \right\rangle} \right]^2
}}{{\epsilon _{{\rm tot}, i}^2 (\lambda )}}},
\end{equation}
where
\begin{equation}
\epsilon _{{\rm tot}, i}^2 (\lambda) \equiv \epsilon _{i}^2
(\lambda) + {\rm rms}^2(\lambda ),
\end{equation}
and $\epsilon _{i}(\lambda)$ is the error in the flux $f_i (\lambda
)$. Here $\left\langle {f(\lambda )} \right\rangle$ is the mean flux while
${\rm rms}(\lambda)$ is the intrinsic scatter -- that is, the rms flux
after removing measurement errors. By maximizing the log-likelihood,
we obtain the mean and rms spectra.  The maximum likelihood method
also provides proper errors in the rms spectra.  We calculated
errors in the inferred mean and rms spectra in a standard way, by
computing their posterior probability distribution after marginalizing
over the other parameters. We adopt $1-\sigma$ errors as symmetric
intervals around the posterior peak containing 68.3\% of the
posterior probability. 

Figure~\ref{fig:compare_rms} compares rms spectra of NGC\,4748
generated with the unweighted rms method, the S/N weighted method,
and the maximum likelihood method, after removing two bad epochs
as described in \S~3.2. As expected, the S/N weighted
rms spectrum is less noisy than the unweighted rms spectrum.
The rms spectrum based on the maximum likelihood method is similar
to but slightly noisier than the S/N weighted spectrum.  
In particular, the maximum likelihood method generates noisy patterns
around the [\ion{O}{3}] line region, 
presumably due to the fact that the error statistics have changed owing to
the subtraction of the strong [\ion{O}{3}] line signals. 
In the case of the mean spectrum, all three methods produce almost identical results.
Thus, we choose the S/N weighting scheme to generate the mean and the
rms spectra, and adopt the errors of the rms spectra from the
maximum likelihood method. 
We note that using the rms spectra 
based on the maximum likelihood method does not significantly 
change the results in the following analysis. 
If more bad epochs with low S/N are removed in
generating rms spectra (as practiced in the reverberation studies;
e.g., Bentz et al. 2009c), the difference among the three methods tends to be smaller.
%However, the S/N weighted and the maximum likelihood methods are preferred 
%from a statistical point of view.
%In principle, including noise in calculations of the rms spectrum is
%preferred from a statistical point of view. 
%However, with the data in hand, we see no
%clear differences in final result when we do so."

We note that there may be a potential bias in the 
S/N weighted method owing to the fact that in the high continuum state
the S/N is higher while emission lines are narrower.
Thus, the S/N weighted rms spectra can be slightly biased toward having 
narrower lines. On the other hand, the time lag between the luminosity change 
and the corresponding velocity change will reduce the bias since
the high luminosity and the corresponding narrow line width are not observed at the
same epoch.

To test this potential bias, we compared the line-width measurements based on 
S/N weighted and unweighted rms spectra, respectively. 
We find that the line width decreases by $2.6\pm2.1$\% for \sigmaline\ 
and $2.7\pm1.3$\% for \FWHM\ when the S/N weighted rms spectra are used, 
indicating that the bias is not significant for our sample AGNs.
However, this offset is not due to the luminosity bias 
since the S/N ratio does not correlate with AGN luminosity.
Instead, the change of the S/N ratio is mostly due to the effects of seeing 
and miscentering within the slit since different amounts of stellar light were 
observed within the slit on different nights. 
Considering the low level of luminosity variability and the time lag, 
the night-to-night seeing and weather variations would be the predominant 
factors affecting the S/N ratio. 
The average offset of $\sim$3\% is dominated by two objects, 
NGC~6814 (0.07 dex for \sigmaline, 0.02 dex for \FWHM) and 
SBS~1116 (0.04 dex for \sigmaline, 0.05 dex for \FWHM), which showed the largest stellar fraction in Fig. 1, 
thus supporting our conclusion. 
By excluding these two objects, the average offset decreases to $\sim$1\%.
Thus, we conclude that the potential AGN luminosity bias in the S/N weighted 
method is not significant, at least for our sample.

\begin{figure*}
\centering
    \includegraphics[width=1.0\textwidth]{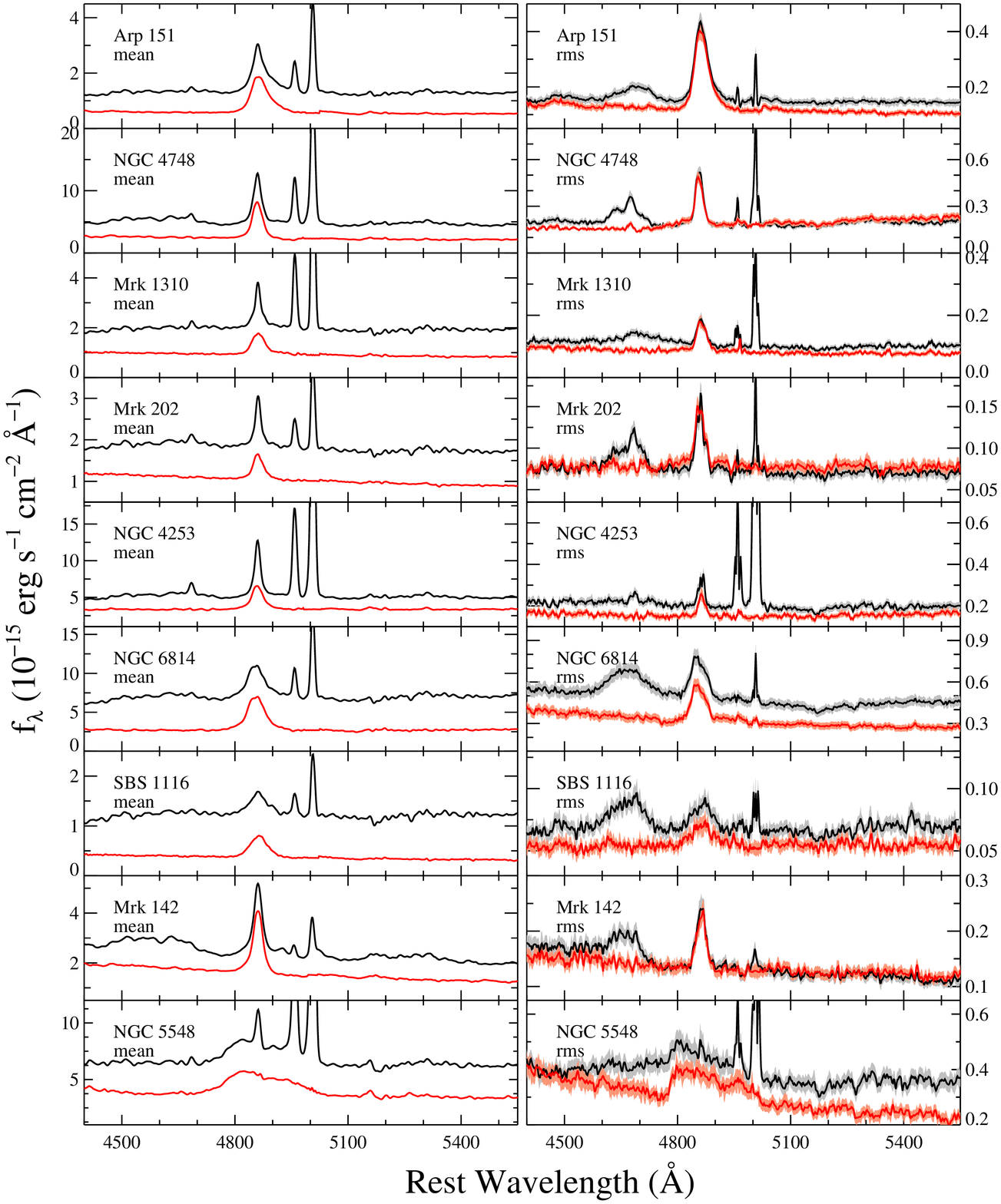}
    \caption{{\it Left:} S/N weighted mean spectra of 9 Seyfert
    galaxies.  {\it Right:} S/N weighted rms spectra.  In each
    panel, red lines represent spectra obtained after removing narrow
    lines, \ion{Fe}{2} emission, \ion{He}{2} lines, and host-galaxy
    starlight from each individual SE spectrum.  Black lines represent
    rms spectra obtained without removing the same components from
    each individual spectrum. Shaded regions show the errors from the 
    maximum likelihood method described in \S 3.3.
    \label{fig:all_mean_rms}}
\end{figure*}

\begin{figure}
\centering
    \includegraphics[width=0.45\textwidth]{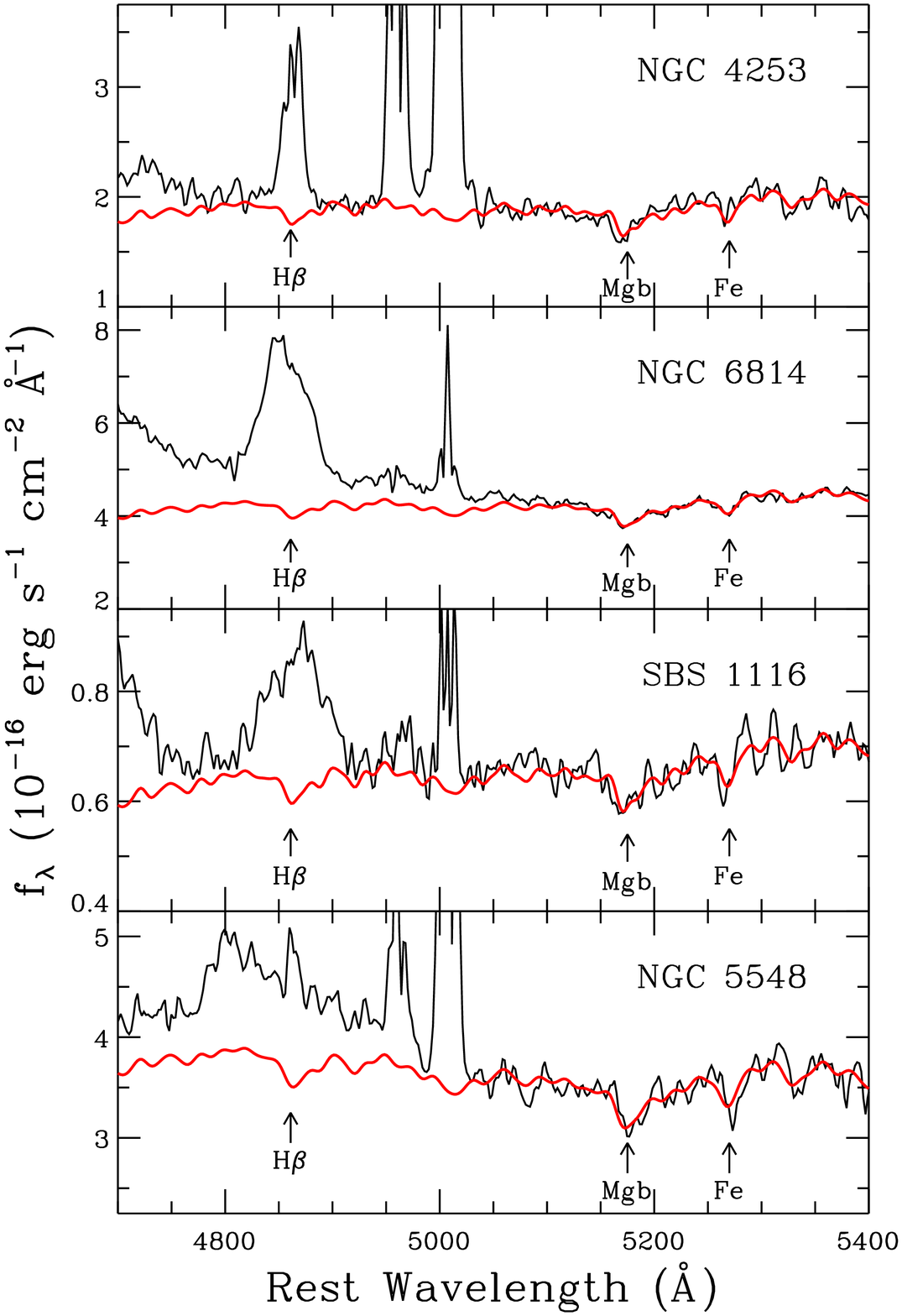}
    \caption{The rms spectra of 4 Seyfert 1 galaxies with strong
     stellar features. Black (red) solid lines represent rms spectra 
     (stellar model fit). Strong absorption lines are labelled with arrows. 
    \label{fig:rms_stellar}}
\end{figure}

\subsubsection{The Effect of Host Galaxy, \ion{Fe}{2}, and \ion{He}{2}}

Although the rms spectra are supposed to contain only varying
components of AGN spectra, residuals of narrow lines (e.g.,
[\ion{O}{3}]) are often present due to residual systematic errors
\citep[due to calibration issues;][]{bentz09c}.  Additionally, the
variation of the host-galaxy starlight contribution to the total flux
can be significant (10--20\%) in the extracted SE spectra as discussed
in \S 3.2.2.  This variable starlight is responsible for the stellar
absorption features often visible in the rms spectra.  To demonstrate
the presence of stellar absorption lines in the rms spectra, we fit
the continuum with a stellar-population model. As shown in Figure \ref{fig:rms_stellar},
it is clear that the rms spectra show stellar absorption lines [such
as the Mg \textit{b} triplet ($\sim5175$~\AA), Fe (5270~\AA), and
possibly \Hb\ (4861~\AA)] for Seyfert 1 galaxies having strong
starlight contribution.  Thus, for AGNs with high starlight fraction,
like the ones considered here, it is important to remove the variable
starlight in order to generate pure AGN rms spectra and correctly
measure the widths of the broad lines.

To minimize these residual features in the rms spectra, we subtracted
the narrow lines in all SE spectra before making mean and rms
spectra. We also subtracted the \ion{Fe}{2} emission blends,
host-galaxy starlight, and the \ion{He}{2} emission line from each SE
spectrum.  In Figure~\ref{fig:all_mean_rms}, we show the S/N weighted
mean and rms spectra with and without prior removal of narrow-line
components, the \ion{Fe}{2} blend, \ion{He}{2} lines, and host-galaxy
starlight.  Clearly, the rms spectra are significantly affected by
this procedure. In particular, removing the \ion{Fe}{2} and
\ion{He}{2} emission changes the continuum shape around \Hb. For the
objects with higher starlight fraction, stellar \Hb\ absorption is
present in the rms spectra, if starlight is not removed from each SE
spectrum.

\begin{figure}
\centering
    \includegraphics[width=0.45\textwidth]{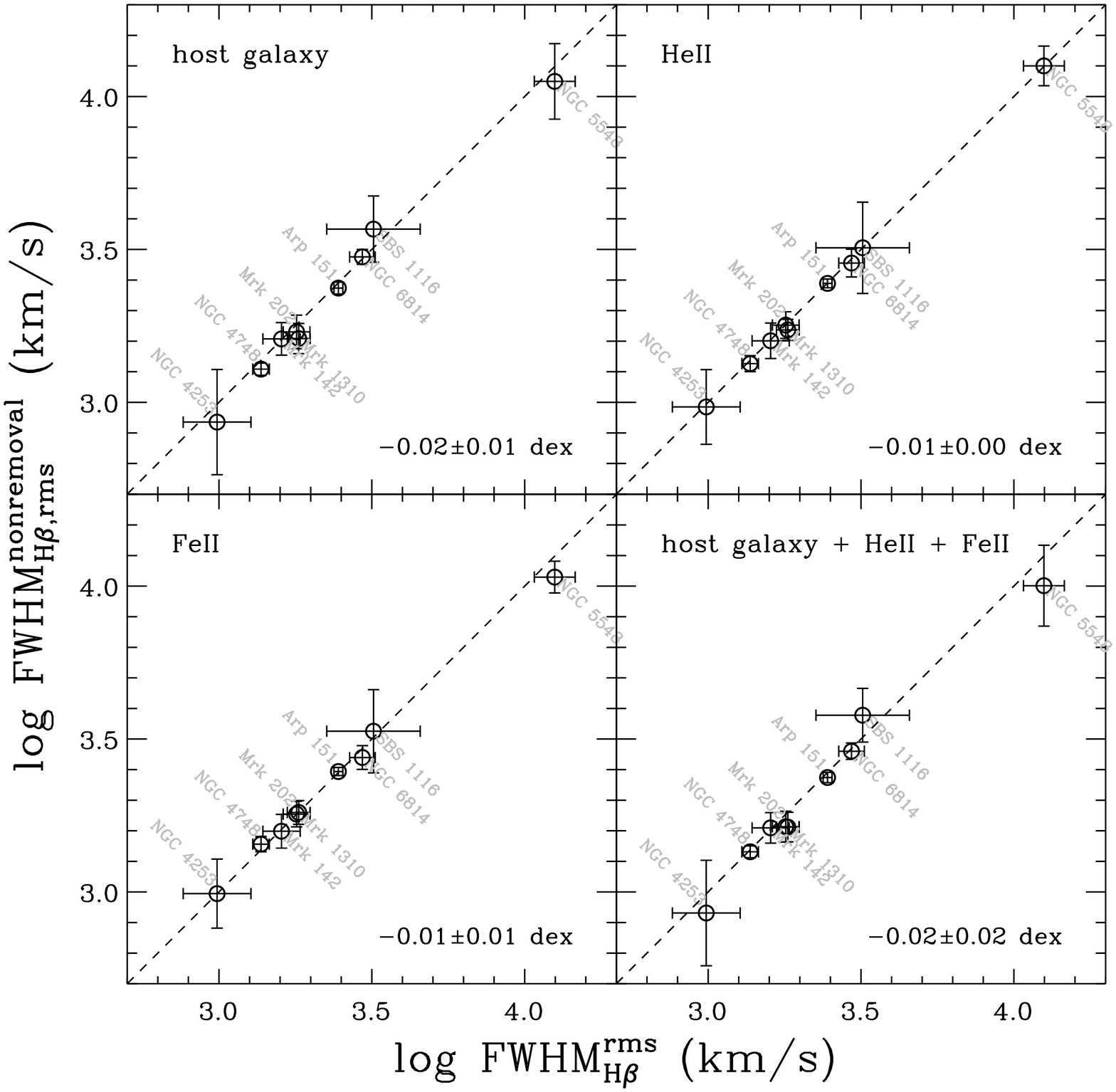}\\
    \includegraphics[width=0.45\textwidth]{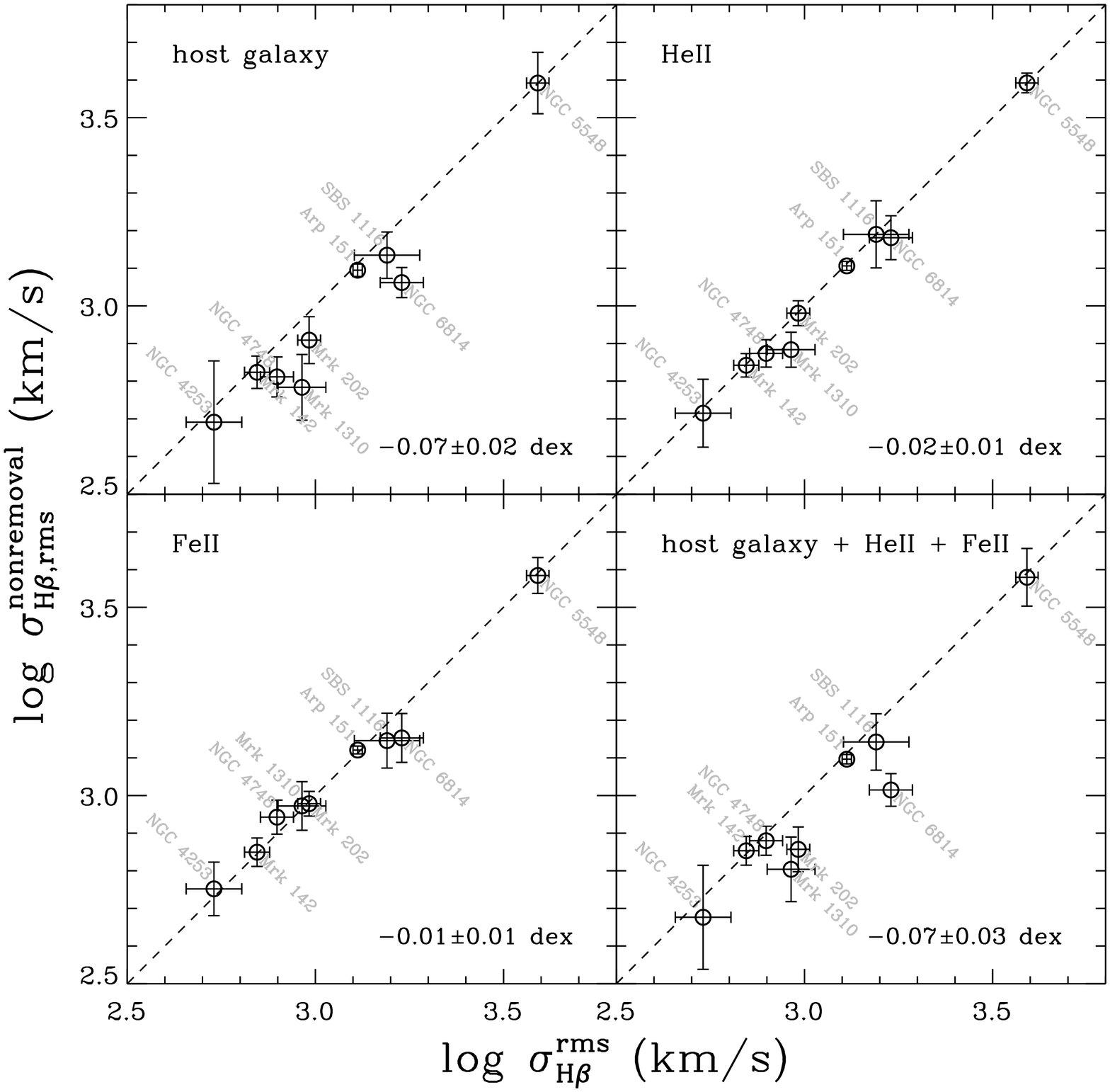}
    \caption{The effect of each blended component on the rms line widths using
    the \FWHM\ (top) and \sigmaline\ (bottom). In each panel, the rms
     line widths without prior removal of each component 
     (given in the upper-left corner) are plotted as a function of final 
     rms line widths (all non-broad-line components removed). 
          The dashed line indicates an identity relationship.
     The average offsets with their standard errors are given in the lower-right 
     corner of each panel. 
    \label{fig:effect_on_rms}}
\end{figure}

To quantify the change of the line widths due to prior removal
of the starlight, \ion{He}{2}, and \ion{Fe}{2} components,
we compared the line-width measurements from rms spectra generated
with/without prior removal.
Each panel in Fig.~\ref{fig:effect_on_rms} shows the effects of individual 
components by comparing the line-width measurements 
from the rms spectra with prior removal of all three components 
(i.e., host-galaxy stellar features, \ion{He}{2}, and the \ion{Fe}{2} blend)
with those from the rms spectra without subtracting one of the three components,
%host-galaxy stellar features, \ion{He}{2}, or the \ion{Fe}{2} blend, 
respectively.
We found that the effect of host-galaxy stellar features is stronger 
than those of \ion{He}{2} and the \ion{Fe}{2} blend. 
Without subtracting host-galaxy stellar features, the rms line widths  
decrease by $18\pm5$\% for \sigmaline~and $4\pm3$\% for \FWHM, 
indicating that the line wings are more affected than the line core. 
The large increase of \sigmaline~can be understood as the buried \Hb~line wings
within the residual of stellar features are restored by subtracting 
the host stellar lines, leading to a lower continuum level 
and larger line width.

The subtraction of \ion{He}{2} changes the line width of objects
that show strong blending with the \Hb\ line (e.g., Mrk~1310 and NGC~6814). 
On average, the effect of \ion{He}{2} on the \Hb\ line width is at the 
$4.7\pm2.2$\% level for \sigmaline~and the $1.6\pm0.6$\% level for \FWHM.
In the case of the \ion{Fe}{2} subtraction, the effect on the rms line widths is 
more complex. 
Line widths increase for some objects and decrease for other objects,
depending on whether the \ion{Fe}{2} emission residual is strong. 
For example, if the \ion{Fe}{2} residual is prominent in the continuum region 
(i.e., 5080--5550\AA), then the removal of \ion{Fe}{2} will lower the
continuum level, increasing the \Hb~line width. 
In contrast, if the \ion{Fe}{2} residual is strong under \Hb,
then the \Hb~line width will decrease by subtracting \ion{Fe}{2}.
On average, the effect of \ion{Fe}{2} on the \Hb~ line width is at the 
$1.2\pm2.9$\% level for \sigmaline~and the $1.6\pm2.2$\% level for \FWHM.

Without prior removal of all three components
(i.e., host-galaxy stellar features, \ion{He}{2}, and the \ion{Fe}{2} blend),
the line widths are underestimated by $18\pm6$\% for \sigmaline~ and $5\pm4$\% 
for \FWHM, due to the combined effects as described above.
Subtracting stellar features has the most significant impact on the 
measurements of rms line dispersion, demonstrating the importance of
prior removal of starlight when stellar contribution is significant.
Moreover, in order to successfully remove the \ion{He}{2} blending
in the rms spectra, the host-galaxy component as well as \ion{Fe}{2} emission blends
should be simultaneously fitted in the modeling of the continuum. 
Thus, we conclude that for AGNs with strong host galaxy starlight,
strong \ion{Fe}{2}, or blended \ion{He}{2}, it is necessary 
to remove all non-broad-line components from SE spectra 
in order to generate the cleanest
rms spectra and reduce errors in measuring the \Hb\ line width. 

\subsubsection{Mean Spectra}

We generated the S/N weighted mean
spectra without prior removal of narrow lines, iron emission, and
host-galaxy starlight.  Then, we used the same multi-component
spectral fitting procedure as used for the SE spectra (see
Fig.~\ref{fig:all_fit_mean}).  
Note that in the case of mean spectra, removing the narrow lines,
\ion{Fe}{2} blends, and host-galaxy absorption features before
or after generating the mean spectra results in almost identical 
\Hb\ broad-line profiles.

\subsubsection{Error Estimation}

Using the S/N weighted rms and mean spectra, we measured the widths
of the H$\beta$ line from the continuum-subtracted spectra and determined
the continuum luminosity at 5100~\AA, as described in \S 3.2.1 and \S
3.2.2.  We estimated the uncertainty of the line-width measurements in
the S/N weighted mean and rms spectra using the bootstrap method
\citep[e.g.,][]{peterson04}.  One thousand samples per object were 
generated. The median and standard deviation of the distribution of
measurements were adopted as our line-width estimate and uncertainty;
these are listed in Table~\ref{table:Hbeta_width}. We also 
estimated the line-width uncertainties for the rms spectra using 
the method given in \S 3.2.3. We found that the errors estimated from
both the Monte Carlo flux randomization and the bootstrapping were 
consistent within a few percent on average, which yielded
almost identical fitting results.

%%%%%%%%%%%%%%%%%%%%%%%%%%%%%%%%%%%%%%%%%%%%%%%%%%%%%%%%%%%%%%%%%%
\begin{deluxetable}{lcccc}
\tablecolumns{5}
\tablewidth{0pt}
\tablecaption{Rest-Frame Broad H$\beta$ Line-Width Measurements}
\tablehead{
\colhead{Object} &
\multicolumn{2}{c}{Mean Spectrum} &
\multicolumn{2}{c}{RMS Spectrum} \\
\cline{2-3}
\cline{4-5}
\colhead{} &
\colhead{\sigmaline} &
\colhead{\FWHM} &
\colhead{\sigmaline} &
\colhead{\FWHM}\\
\colhead{} &
\colhead{(km s$^{-1}$)} &
\colhead{(km s$^{-1}$)} &
\colhead{(km s$^{-1}$)} &
\colhead{(km s$^{-1}$)}
}
\startdata
Arp\,151        &  $1726\pm  17$  &  $3076 \pm  39$ &  $1295\pm  37$ & $2458\pm   82$ \\
NGC\,4748       &  $952 \pm   6$  &  $1796 \pm   8$ &  $ 791\pm  80$ & $1373\pm   86$ \\
Mrk\,1310       &  $1229\pm  12$  &  $2425 \pm  19$ &  $ 921\pm 135$ & $1823\pm  157$ \\
Mrk\,202        &  $1047\pm   8$  &  $1787 \pm  15$ &  $ 962\pm  67$ & $1794\pm  181$ \\
NGC\,4253       &  $1232\pm   9$  &  $1946 \pm  10$ &  $ 538\pm  92$ & $ 986\pm  251$ \\
NGC\,6814       &  $1744\pm  12$  &  $3129 \pm  14$ &  $1697\pm 224$ & $2945\pm  283$ \\
SBS\,1116+583A  &  $1460\pm  23$  &  $3135 \pm  36$ &  $1550\pm 310$ & $3202\pm 1127$ \\
Mrk\,142        &  $970 \pm   5$  &  $1671 \pm   6$ &  $ 700\pm  54$ & $1601\pm  224$ \\
NGC\,5548       &  $4354\pm  25$  &  $12402\pm 111$ &  $3900\pm 266$ & $12539\pm1927$
\enddata
\label{table:Hbeta_width}
\end{deluxetable}
%%%%%%%%%%%%%%%%%%%%%%%%%%%%%%%%%%%%%%%%%%%%%%%%%%%%%%%%%%%%%%%%%%%%%%%%%%%
% Analysis and Results
%%%%%%%%%%%%%%%%%%%%%%%%%%%%%%%%%%%%%%%%%%%%%%%%%%%%%%%%%%%%%%%%%%%%%%%%%%%

\section{ANALYSIS AND RESULTS} \label{section:analresul}

\subsection{Testing the Assumptions of SE BH Mass Estimators}

Single-epoch \mbh\ estimates are based on the ``virial'' assumption
and on the empirical relation between BLR size and AGN
luminosity. Since \mbh\ does not vary over the time scale of our
campaign, AGN luminosity and line velocity should obey the relation 
$V^2 \propto L^{-0.5}$.  In this section, we test this assumption by
studying the relation between the line width and continuum luminosity
from individual SE spectra of Arp~151, the object with the
highest variability during the LAMP campaign.

\begin{figure}
\centering
    \includegraphics[width=0.45\textwidth]{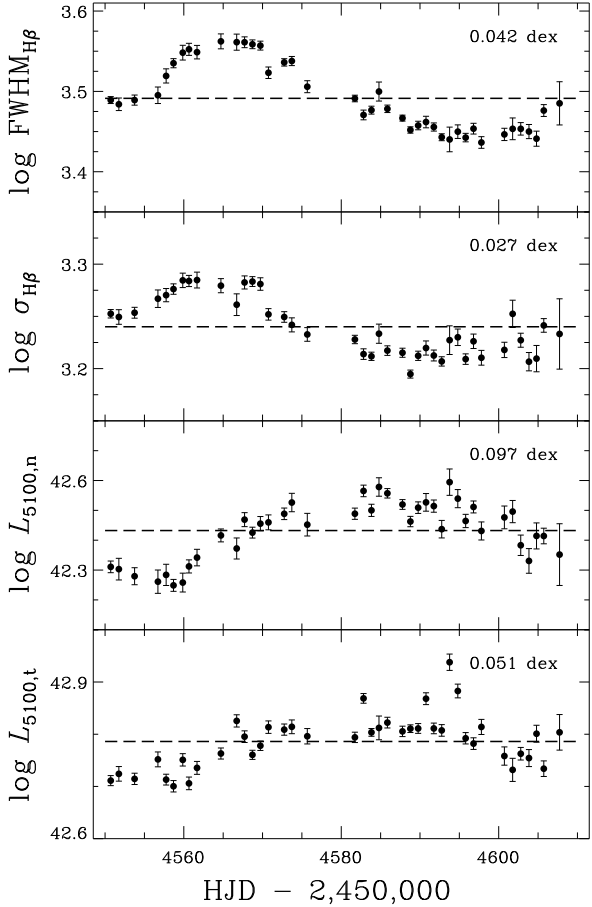} 
    \caption{
    Time variations of the \Hb\ line width ({\it top:} \FWHM; {\it
    upper middle:} \sigmaline) and the continuum luminosity at
    5100~\AA\ ({\it lower middle:} nuclear; {\it bottom:} total) of Arp
    151.  The dashed lines represent the average values over the
    monitoring period.  The rms dispersion values are given in each
    panel.
    \label{fig:arp151_variation}}
\end{figure}

In Figure~\ref{fig:arp151_variation} we present the time variation of
the line width and luminosity of Arp~151. Line width and luminosity
are inversely correlated, although the variability amplitude is
smaller in luminosity than in line width. Ideally, the luminosity
variability should be four times as large as the line-width
variability ($0.042\pm0.001$ dex and $0.027\pm0.003$ dex,
respectively, for \FWHM\ and \sigmaline).  However, one must take into
account the residual contamination from nonvariable sources to the
observed continuum. In fact, the amplitude of the luminosity
variability is significantly smaller than expected based on the
line-width variability if the total luminosity is used (bottom
panel). In contrast, providing a validation of our constant continuum
subtraction procedure, the variability amplitude of the nuclear
continuum is consistent with that expected from the line width, as we
will further quantify below.

\begin{figure}
\centering
    \includegraphics[width=0.45\textwidth]{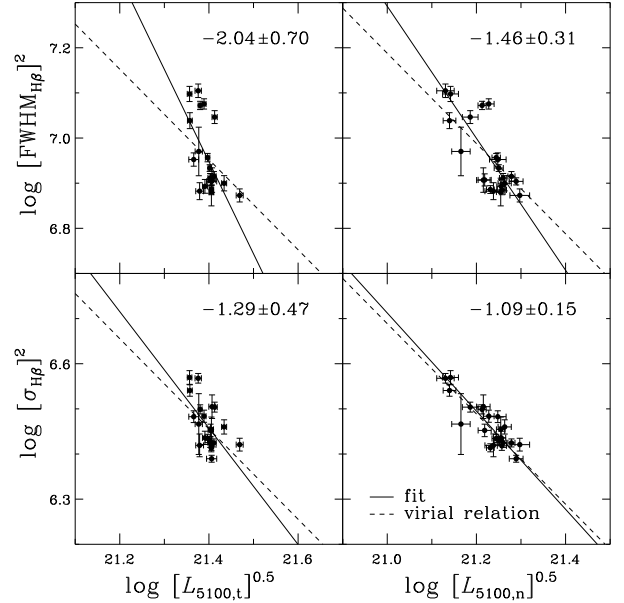}
    \caption{Test of SE mass estimates for Arp 151. Each
    filled circle represents SE measurements after shifting
    luminosity measurements by the average lag of 4 days ({\it left:}
    total luminosity; {\it right:} nuclear luminosity).  Dashed lines
    represent the correlation $L^{0.5} \propto V^{-2}$ expected from
    the virial theorem and the size-luminosity relation, while solid
    lines are the best-fit slopes. The values of best-fit slopes and 
    its uncertainties are given in each panel.
    \label{fig:arp151_virial}}
\end{figure}

In Figure~\ref{fig:arp151_virial}, we compare measured luminosities
and line widths in order to test whether they obey the expected
relation $V^2\propto L^{-0.5}$. The continuum variation is shifted by
the measured time lag, 4 days, to account for the time delay between
the central engine and BLR variations and then matched with the
corresponding epochs of line-width variations. Note that densely
sampled light curves are required for this correction.  As expected,
the observed correlation between total flux $L_{\rm 5100,t}$ and line
width is steeper than the theoretical correlation.  In
contrast, the correlation between nuclear flux and line width is
consistent with the theoretical expectation. 
%And the \FWHM, shows steeper slope than \sigmaline, implying that 
%the \sigmaline~is less biased velocity measure than \FWHM. 
The best-fit slopes\footnote{We used the Bayesian linear regression routine
\texttt{linmix\_err} developed by \citet{kelly03} in the NASA IDL
 Astronomy User's Library. This method is currently the most
 sophisticated regression technique, which takes into account
 intrinsic scatter and nondetections as well as the measurement errors
 in both axes, generating the random draws from posterior probability
 distribution of each parameter for the given data using MCMC
 sampling. In this study, we take best-fit values and uncertainties of
 parameters as the median values and $\pm1\sigma$ standard deviation
 of $10,000$ random draws from corresponding posterior distributions.}
 are $-1.46\pm0.31$ (with intrinsic scatter $0.05\pm0.01$ dex) for
 \FWHM,\ and $-1.09\pm0.15$ (with intrinsic scatter
 $0.02\pm0.01$ dex) for \sigmaline, which is consistent with the
 expected value of $-1$.  The linear correlation coefficients between
 the nuclear luminosity and the line widths are $-0.86$ for the line
 dispersion and $-0.77$ for the FWHM, indicating the tighter inverse
 correlation of continuum luminosity with the line dispersion than
 with the FWHM.

The agreement of the observed correlations with those expected for an
ideal system is remarkable, considering the many sources of noise in
the observed velocity-luminosity relation. 
They include residual errors in the subtraction of the host-galaxy starlight contribution
and the measurement uncertainties of line widths and luminosities.
The inverse correlation between line width
and luminosity further corroborates the use of SE mass estimates
\citep{PW99,PW00,kolla03,peterson04}.

%%%%%%%%%%%%%%%%%%%%%%
% line widths (sigma,FWHM)
%%%%%%%%%%%%%%%%%%%%%%
\subsection{Uncertainties Due to Variability}

Since the line width and continuum luminosity of an AGN vary as a
function of time, mass estimates from SE spectra may also vary. Owing
to its stochastic nature, this variability can be considered a source
of random error in SE mass estimates. In this section, we quantify
this effect by comparing SE measurements with measurements from the
mean spectra.

\subsubsection{The Effect of Line-Width Variability}

\begin{figure}
\centering
    \includegraphics[width=0.45\textwidth]{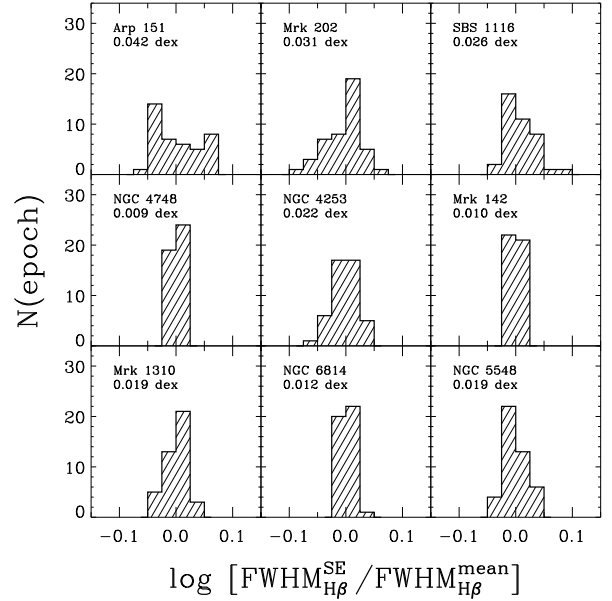}
    \caption{Distribution of the \FWHM\ measured from all
    SE spectra.  Each \FWHM\ value is normalized to the
    \FWHM\ measured from the mean spectra. The average
    rms dispersion of $9$ objects is $0.021\pm0.004$ dex.
    \label{fig:histo_FWHM}}
\end{figure}
\begin{figure}
\centering
    \includegraphics[width=0.45\textwidth]{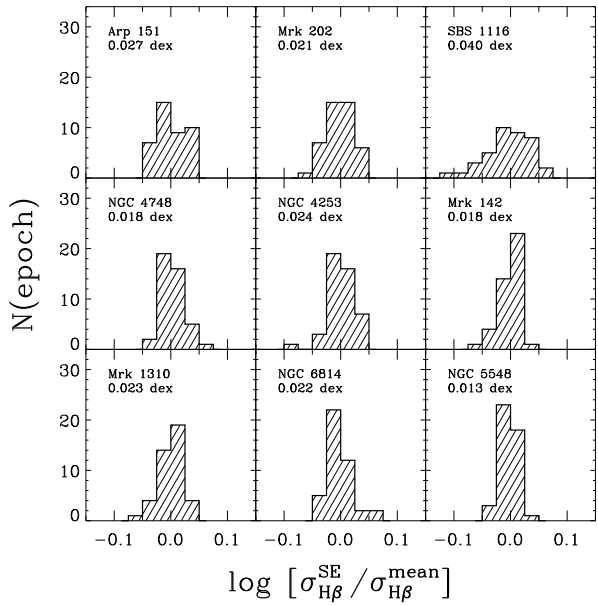}
    \caption{Same as in Fig.~\ref{fig:histo_FWHM}, but for
    \sigmaline.  The average rms dispersion of $9$ objects
    is $0.023\pm0.003$ dex.  \label{fig:histo_sigma}}
\end{figure}

We quantify the dispersion of the distribution of line-width
measurements using all SE spectra. This dispersion can be interpreted
as a random error due to the combined effect of variability and
measurement errors.  In Figure~\ref{fig:histo_FWHM}, we present the
distributions of \FWHM\ measurements from all SE spectra, after
normalizing them by the measurement from the mean spectra. All SE
values are normalized to the FWHM measured from the mean spectra.  The
standard deviation of the FWHM distributions ranges from $0.009$ dex
to $0.042$ dex, with an average of $0.021\pm0.004$ dex ($\sim$5\%)
across all objects.  Note that the standard deviation includes the
variability and the measurement error.

In Figure~\ref{fig:histo_sigma}, we plot the distributions of line
dispersion for all objects. The dispersion of distributions ranges
from $0.013$ dex to $0.040$ dex, with an average and rms of
$0.023\pm0.003$ dex ($\sim$5\%) for the entire sample.  SBS~1116 shows
the broadest distribution; however, part of this scatter can be
attributed to the residual systematic in the left wing of \Hb\ due to
the bad pixels in the original spectra, as discussed previously.

By averaging the standard deviation of the distribution of the
line-width measurements for all 9 objects in the sample, we find that
the uncertainty of SE BH mass estimates due to the line-width
variation and measurement errors is on average $0.044$ dex.  Note that
the dispersion of the line-width distribution strongly depends on the
variability. For example, Arp~151 has the largest variability
amplitude and also the largest variability in the line width. This is
expected if line flux correlates with BLR size and both are connected
to the BH mass. Based on these results, we conclude that the
typical uncertainty of SE mass estimates due to line-width variability
is $\sim$10\%. However, as discussed below, this uncertainty is partly
cancelled out in the virial product by the inverse correlation with
the variability of the continuum.
%%%%%%%%%%%%%%%%%%%%%%%%%%%%%%%%%%%%%%%%%%%%%%%%%%%%%%%%%%%%%%%%%%

\subsubsection{The Effect of Luminosity Variability}

\begin{figure}
\centering
    \includegraphics[width=0.45\textwidth]{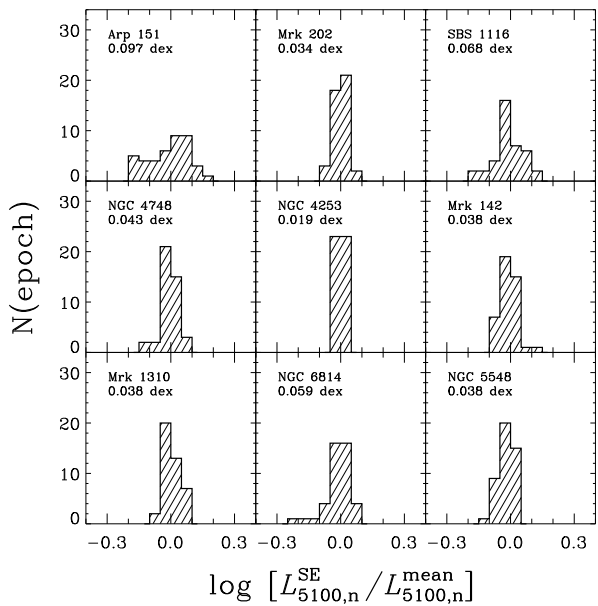}
    \caption{Distribution of nuclear luminosities measured from
    single-epoch spectra (see \S \ref{section:meas:conti}).  Each
    luminosity value is normalized to the nuclear luminosity measured
    from the mean spectra.  The standard deviation of the distribution
    is given in each panel. The average standard deviation of all $9$
    objects is $0.048\pm0.008$ dex.  \label{fig:histo_nuclear_L}}
\end{figure}

We now consider the effect of luminosity variability on SE mass
estimates.  In Figure~\ref{fig:histo_nuclear_L}, we present the
distributions of the nuclear luminosities at 5100~\AA, after
normalizing them by the nuclear luminosity measured from the mean
spectra.  
The standard deviation of the luminosity distributions ranges from
$0.019$ to $0.097$ dex, with an average of $0.048\pm0.008$ dex
($\sim$12\%), which can be treated as a random error of the
continuum luminosity measured from a SE spectrum due to
the luminosity variability and measurement error. 

Based on the empirical size-luminosity relation, the random errors of
the luminosity enter the uncertainty of the SE mass estimates as the
square root (i.e., $0.024$ dex).  This is somewhat smaller than the
uncertainty of SE mass estimates due to the line-width variability,
$0.044$ dex, as determined in the previous section, indicating that
the two do not cancel each other exactly.

\subsubsection{Combined Effect}

Since the luminosity and the line width are inversely correlated as
$V^{2} \propto L^{-0.5}$, one may expect that the variability of
luminosity and line width cancel out in the SE mass
estimates. However, the two effects may not compensate each other
exactly, for a variety of reasons. First, there is a time lag between
continuum and emission-line variability. Second, variations such as in
the ionizing flux may indicate that the luminosity at 5100~\AA\ traces
the broad-line size only approximately. In order to quantify the
combined effect of the continuum luminosity and line-width
variability, we thus investigate the distribution of the virial
product $L_{5100\AA,n}^{0.5} \times\sigma_{{\rm H}\beta}^{2}$ as measured
from SE spectra.

\begin{figure}
\centering
    \includegraphics[width=0.45\textwidth]{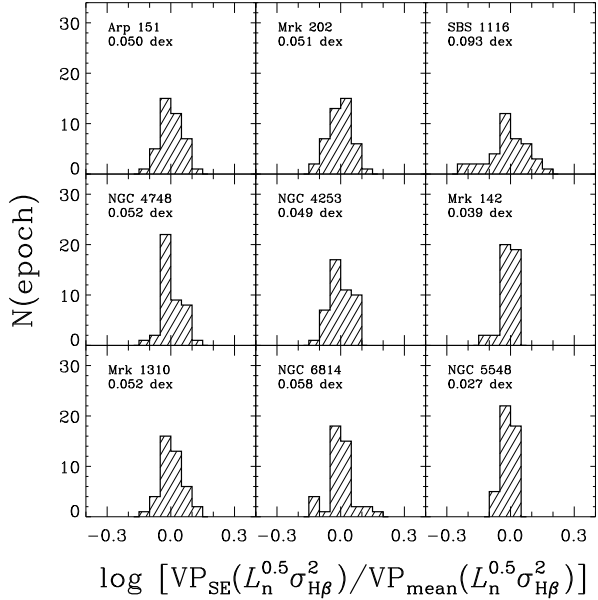}
    \caption{Distribution of the SE virial product ($V^2
    \times L^{0.5}$) normalized to that of the mean spectra. \Hb\ line
    dispersion is used for the velocity and the nuclear luminosity at
    5100~\AA, corrected for the host galaxy, is used for the
    luminosity. The average rms dispersion of all $9$ objects is
%$0.049\pm0.006$ dex ($0.052\pm0.006$ dex) for \FWHM~(\sigmaline).
    $0.052\pm0.006$ dex.
    \label{fig:histo_combined_sigma_nuclear}}
\end{figure}

In Figure~\ref{fig:histo_combined_sigma_nuclear}, we present the
distribution of the SE virial products, normalized by the virial
product measured from the mean spectra.  The standard deviation of the
distributions can be treated as a random error due to the combined
variability and measurement errors.
The average rms scatter (corresponding to a source of random
measurement errors when using the SE estimator) of the virial products
is $0.052\pm0.006$ dex when the line dispersion
($\sigma_{H\beta}$) is used, and $0.049\pm0.006$ dex when
\FWHM\ is used.  

In agreement with previous studies \citep{wilhite07, woo07, denney09},
these results suggest that BH masses based on SE spectra taken at
different epochs are consistent within $\sim0.05$ dex ($\sim 12$\%)
uncertainty, negligible with respect to other sources of uncertainty
which are believed to add up to $\sim$0.4--0.5 dex (see \S5.1).

\subsection{Systematic Difference between SE and Reverberation Masses}

In order to assess the accuracy of the SE mass estimates, we need to
compare the SE masses with the masses determined from reverberation
mapping. Setting aside potential differences in the virial
coefficient, there are two main sources of systematic uncertainties in
SE mass estimates.  One is the potential difference of the line
profile between SE spectra and the rms spectra.  The other is the
systematic uncertainty of the size-luminosity relation.  We postpone
discussion of the latter to a future paper when more accurate
{\it HST}-based nuclear luminosities will be available. Therefore, in this
section we focus on the systematic difference of the \Hb\ line profile
and derive new SE mass estimators recalibrated to account for the
difference found.

%%%%%%%%%%%%%%%%%%%%%%
% line width comparison
%%%%%%%%%%%%%%%%%%%%%%
\begin{figure}
\centering
    \includegraphics[width=0.45\textwidth]{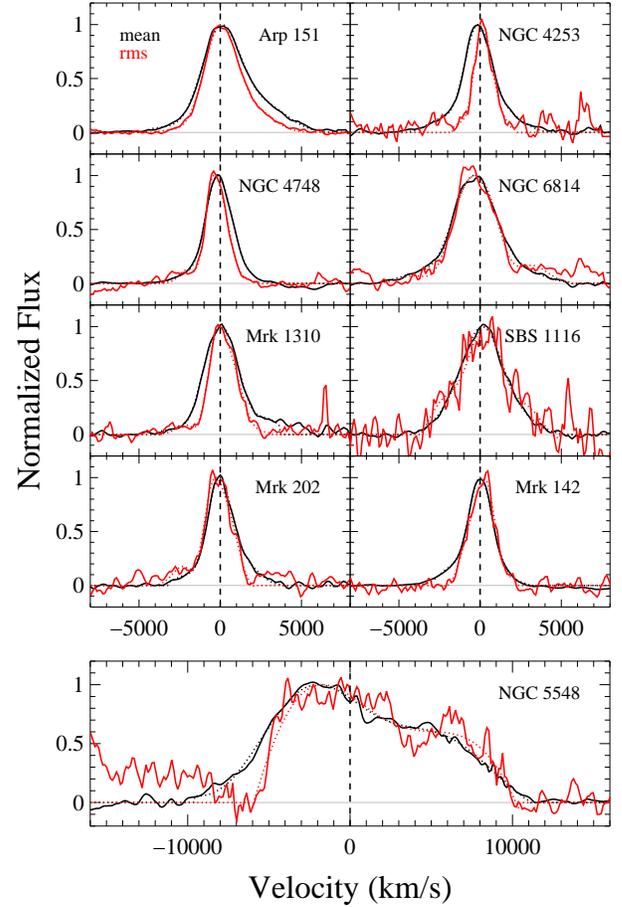}
    \caption{Comparison of the H$\beta$ broad-line profiles in the
    mean (black) and rms (red) spectra. In each panel, solid lines
    represent the data, while dotted lines represent the Gauss-Hermite
    series fitting results.  Each line profile is normalized by
    the maximum value of the fit.  Dashed vertical lines indicate the
    center of the H$\beta$ line.
    \label{fig:compare_spec_mean_rms_data_fits}}
\end{figure}
%%%%%%%%%%%%%%%%%%%%%%
% velocity offset
%%%%%%%%%%%%%%%%%%%%%%
%
\begin{figure*}
\centering
    \includegraphics[width=0.4\textwidth]{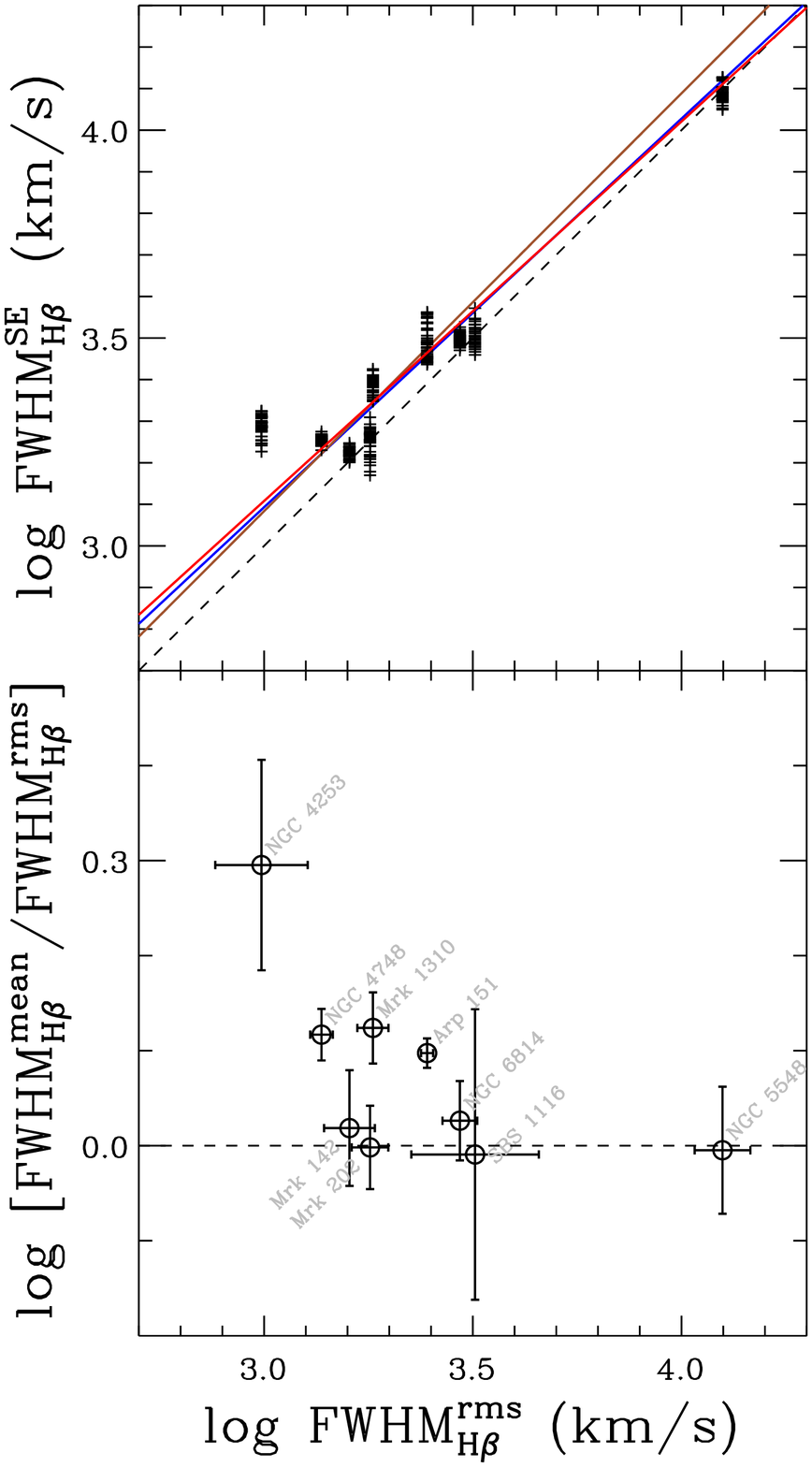} \hspace{2em}
    \includegraphics[width=0.4\textwidth]{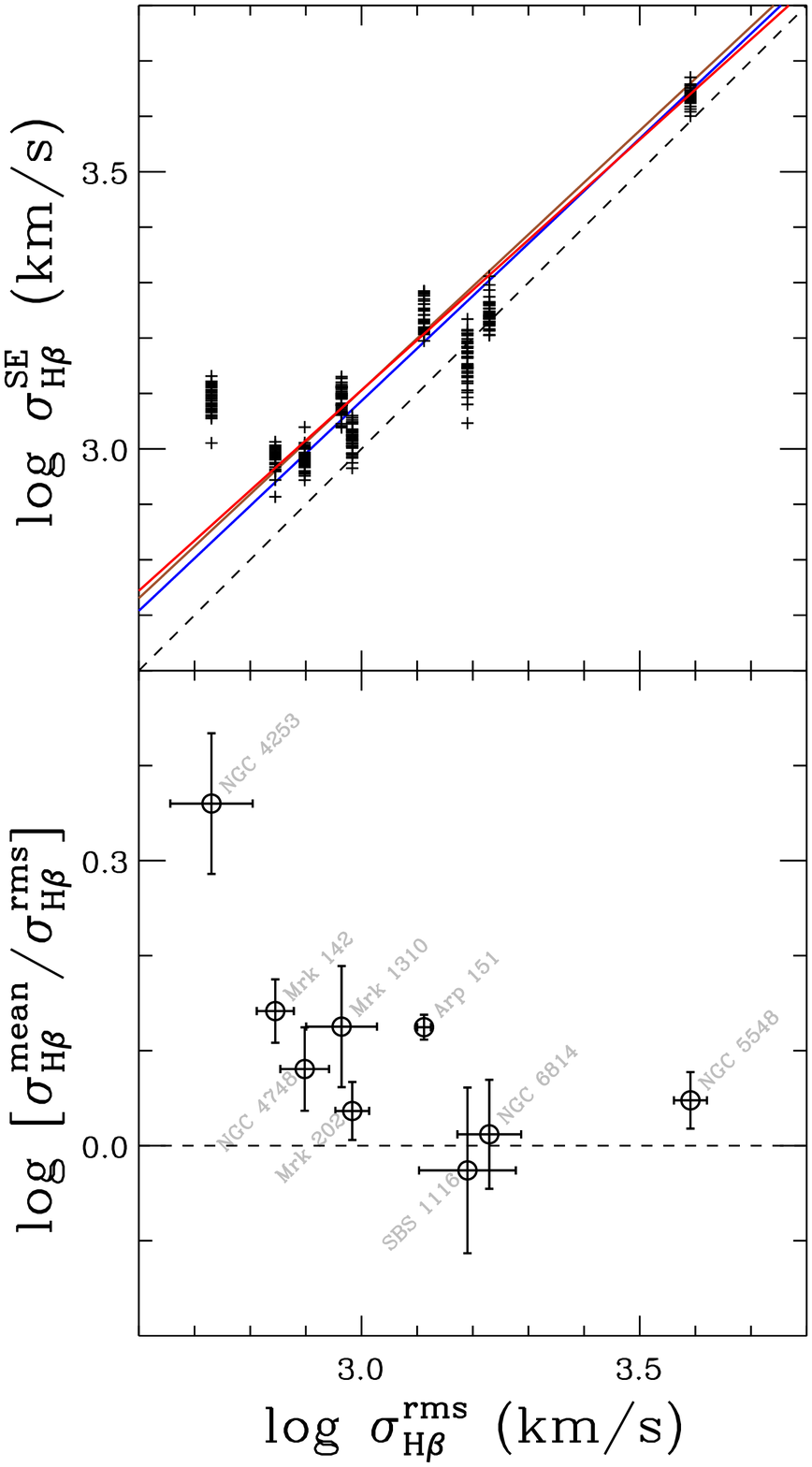}
    \caption{Direct comparisons of line widths and their ratios, i.e., \FWHM\ (left) and
    \sigmaline\ (right), measured from SE (or mean) and rms spectra, as
    a function of line width. Average offset is $0.07\pm0.03$ dex
%($0.05\pm0.02$ dex excluding NGC\,4253) 
   for \FWHM\ and $0.10\pm0.04$ dex
%($0.07\pm0.02$ dex excluding NGC\,4253) 
    for \sigmaline. 
%20\% fixed error-bars are shown in bottom-right corner on each top panel.
    Dashed lines indicate an identity relation while solid lines are 
    the best-fit results using all objects (red), excluding NGC\,4253 (blue),
or excluding NGC\,5548 (brown) using bootstrap errors.
    \label{fig:offset_vel}}
\end{figure*}
%%%%%%%%%%%%%%%%%%%%%%
% VP offset
%%%%%%%%%%%%%%%%%%%%%%
\begin{figure}
\centering
    \includegraphics[width=0.4\textwidth]{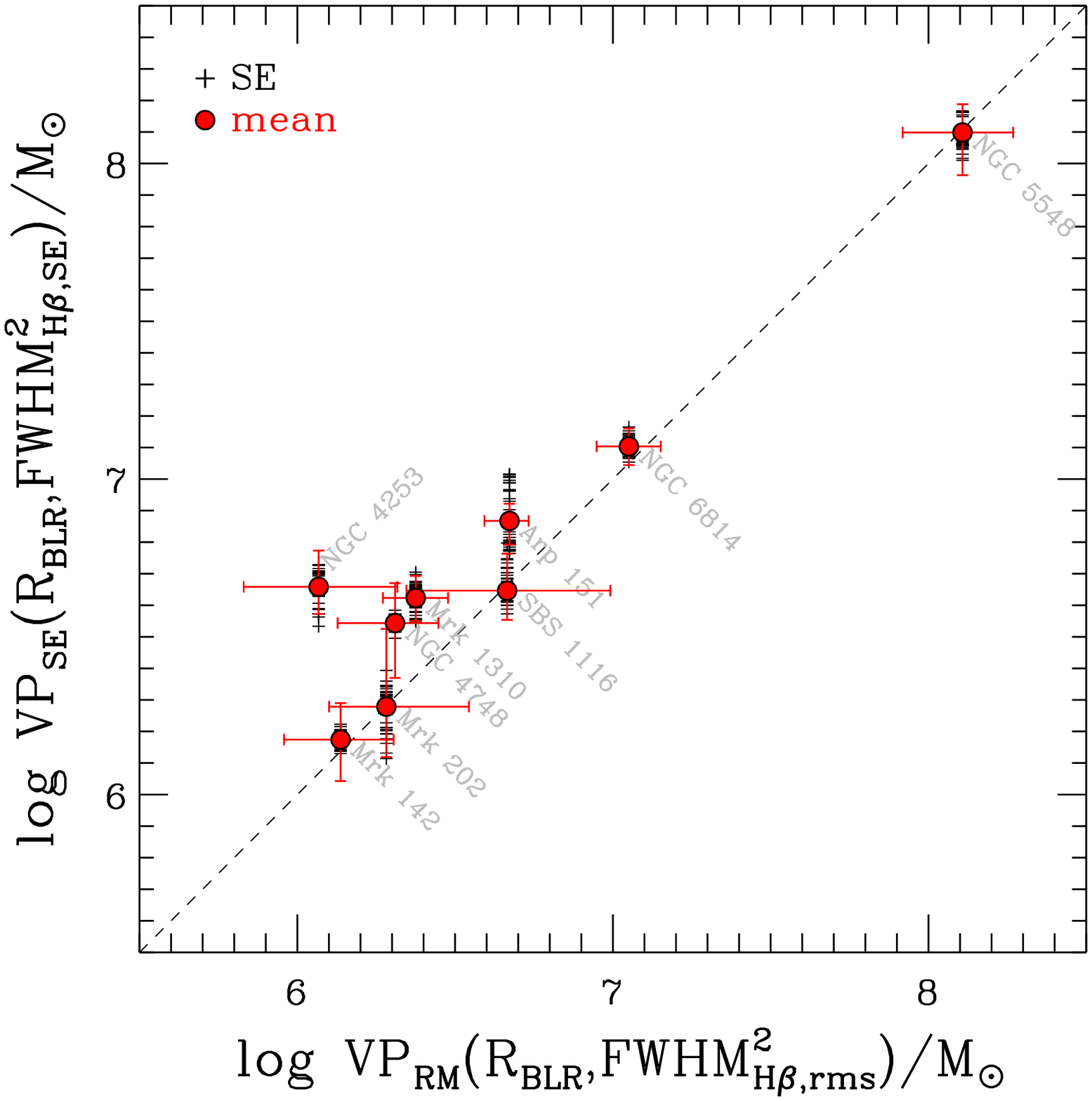}
    \includegraphics[width=0.4\textwidth]{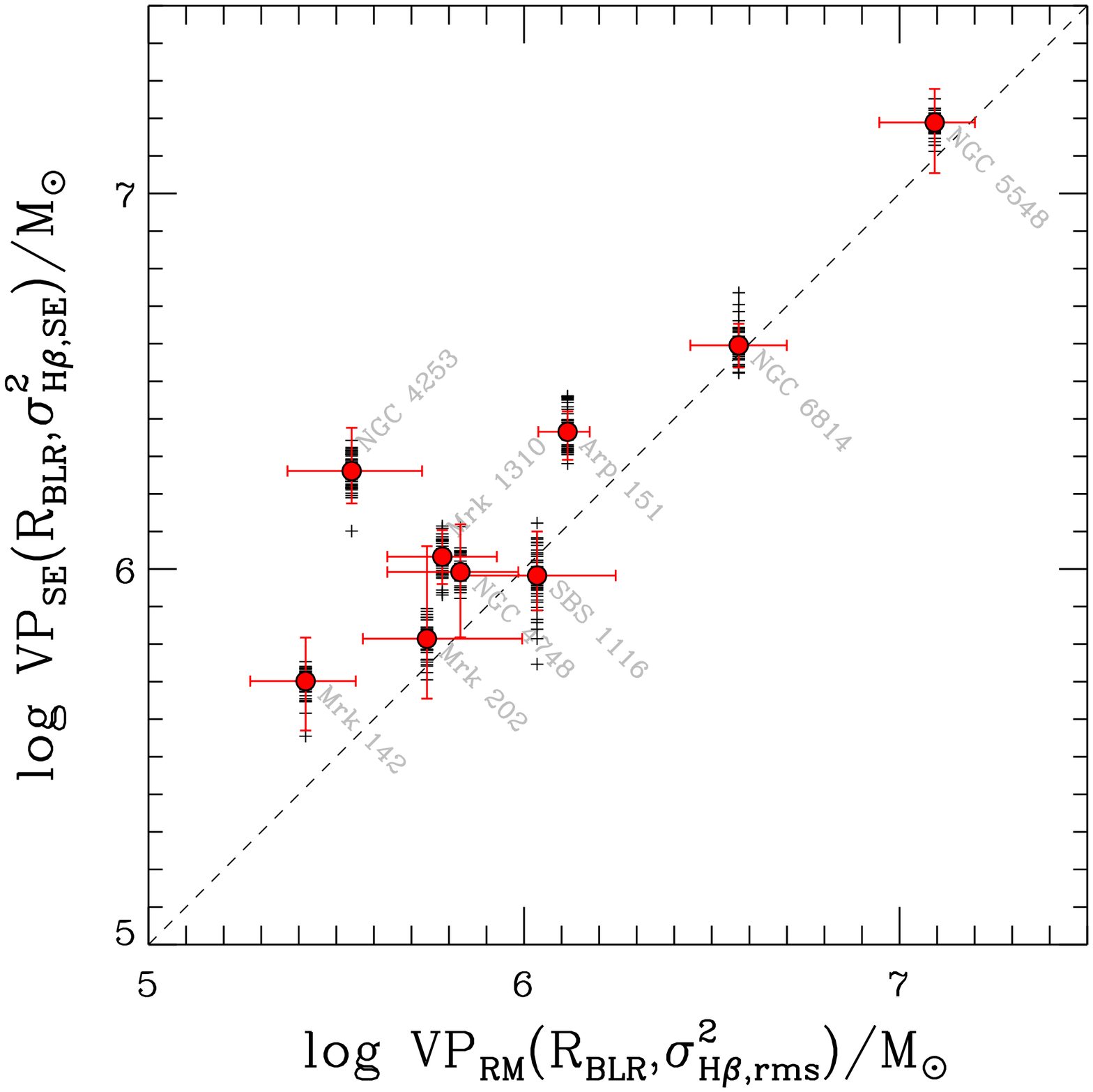}\\
    \caption{Comparison of the virial products measured from SE
    spectra and rms spectra.  Virial products are calculated using
    \FWHM\ (top) or \sigmaline\ (bottom).  In addition to measurements
    from each SE spectrum (crosses), we show those from the
    mean spectra (filled circles).  
    %Multiple measurements of NGC\,5548
    %(blue circles) are taken from Collin et al. (2006).  % 
    %The
    %average offset of all SE measurements is $0.39\pm0.38$
    %($0.26\pm0.38$) dex %when \sigmaline~ (\FWHM) is used for the
    %virial products.  % 
    %
    %including NGC\,5548 from Collin et al. (2006) 
    The average offset of all mean measurements is $0.201\pm0.075$
    ($0.147\pm0.066$) dex when \sigmaline\ (\FWHM) is used for the
    virial products.  \label{fig:offset_mass}}
\end{figure}

\subsubsection{Comparing Line Profiles}

In Figure~\ref{fig:compare_spec_mean_rms_data_fits} we compare the broad
\Hb\ line profiles measured from the mean and rms spectra 
after normalizing by the peak flux.  Generally the \Hb\ line is broader
in the mean spectra than in the rms spectra, indicating that the
variation is weaker in the line wings than in the line core.  It is
worth noting that the observed offset cannot be explained by the
contamination of the narrow \Hb\ component or the \ion{Fe}{2} blends
since we consistently subtracted them in both the mean and rms
spectra.  To verify this we arbitrarily decreased the amount of narrow
component subtracted from the observed H$\beta$ profile, and found
that the large offsets between rms and mean spectra are virtually
unchanged.

The broader line width in the mean spectra has been noted in previous
reverberation studies (e.g., Sergeev et al. 1999; Shapovalova et
al. 2004). \citet{collin06} reported that the line widths in the mean
spectra were typically broader by $\sim 20$\% than those in the rms
spectra. \citet{denney10} also found that some objects in their
reverberation sample clearly showed narrower line widths in the rms
spectra than in the mean spectra. Several different and somewhat
mutually exclusive explanations have been suggested for this
difference.  For example, \citet{shields10} explained the systematic
difference of the line width as being due to the high-velocity gas in
the inner BLR being optically thin to the ionizing continuum and hence
fully ionized. In this way, the line wings have weak variability and
are suppressed in the rms spectra.  In contrast,
\citet{KG04} suggested a distance-dependent responsivity of optically
thick clouds to explain the weak variability of Balmer line wings.

We quantify the systematic offset in line width in
Figure~\ref{fig:offset_vel} by showing the ratios of the line width
measured from the mean (and SE) spectra to those measured from the rms
spectra as a function of line width.  The average offset in
\FWHM\ is $0.07\pm0.03$ dex ($0.05\pm0.02$ dex, if NGC\,4253, the object 
with the narrowest line, is excluded).  In the case of line dispersion
(\sigmaline), the offset is slightly larger, $0.10\pm0.04$ dex
($0.07\pm0.02$ dex if NGC\,4253 is excluded).  The larger offset of the
line dispersion in comparison with FWHM is consistent with there being 
mainly a difference between variability in the wings and in the core.

There seems to be a systematic trend, in the sense that the offset
becomes relatively larger for the narrower line objects, but its
origin is not clear. In particular, the narrow-line 
Seyfert 1 galaxy NGC\,4253 (Mrk\,766) has the narrowest \Hb\
line in the sample and shows the largest
systematic difference.  It is possible that the systematic difference
for this particular object with very narrow \Hb\ (\FWHM\ (rms) $<1000$
\kms) may be amplified due to imperfect subtraction of the narrow component, 
the \ion{Fe}{2} blends, or starlight. However, the trend
is present even if we remove this object from the sample.

In order to correct for this potential bias, we derive a relation
between H$\beta$ line width as measured from rms and SE spectra by
fitting the trend as shown in Figure~\ref{fig:offset_vel}.
Using the linear regression routine \texttt{linmix\_err} \citep{kelly03}, 
we fit the linear relationship in log-scale using bootstrap errors
determined in \S 3.3.
We also determined the slope excluding the narrowest-line object (NGC\,4253)
or the broadest-line object (NGC\,5548) from the sample.
As shown in Figure 13, removing either NGC\,4253 or NGC\,5548 
from the sample does not significantly change the slope. 

In addition, we fit the slope using a fixed error for all objects.
Since the bootstrap errors on the rms line widths are significantly 
different for each object owing to the different quality and S/N ratios 
of individual single-epoch spectra, we assigned a fixed error, such as
20\% on both axes, to test the effect of errors.
The best-fit slope using a fixed error
is slightly shallower than that with bootstrap errors
since the most offset object, NGC\,4253 has a large bootstrap 
error and consequently has smaller weight in the fitting process. 

To secure a large dynamic range, we decided to use all 9 objects for the fit
and adopt the best-fit result using bootstrap errors.
The adopted best fits are expressed as
%
%log FWHM({\rm rms}) = -0.40460306(+/-0.050791750) + 1.0954480(+/-0.014792599) log FWHM({\rm SE})   (I.S. 0.040349556+/-0.003043265)
\begin{equation}
\begin{split}
&\log {\rm FWHM}_{\rm H\beta}({\rm rms}) = \\
  &- 0.405 (\pm 0.051) + 1.095 (\pm 0.015)~\log {\rm FWHM}_{\rm H\beta}({\rm SE}),
\end{split}
\label{eqn:convert_velFWHM}
\end{equation}
%log SIGMA(rms) = -0.43445812(+/-0.060041827) + 1.1058353(+/-0.018852891) log SIGMA(SE)   (I.S.  0.062630835+/-0.0047458204)
%
\begin{equation}
 \log \sigma_{\rm H\beta}(rms)  =  - 0.434 (\pm 0.060) + 1.106 (\pm 0.019)~\log \sigma_{\rm H\beta}(SE).
\label{eqn:convert_velsigma}
\end{equation}
\noindent
However, these fits should not be extrapolated to
high-velocity objects; otherwise, negative bias will be introduced. 
Since our sample consists of relatively narrow-line Seyfert 1 galaxies,
we recommend that readers use Eqs. \ref{eqn:convert_velFWHM} and \ref{eqn:convert_velsigma} 
for objects with FWHM$_{\rm H\beta, SE}< 3,000$ \kms\ and $\sigma_{\rm H\beta, SE}<2,000$ \kms, respectively.  

\subsubsection{Systematic Offset of Mass Estimates}

The systematically broader line width in SE spectra would result in
overestimates of SE masses if unaccounted for.  In
Figure~\ref{fig:offset_mass} we plot the ratio of the SE virial product
(VP) with respect to the virial product based on the reverberation
studies as a function of the virial product.
Note that to demonstrate the effect of the systematic difference between
SE and rms spectra, we simply used the measured $R_{\rm BLR}$ for all SE virial products,
instead of using $L_{5100}$ and the size-luminosity relation. 
As expected, virial products exhibit a biased systematic
trend.  The average offset of all SE measurements is
$0.152\pm0.009$ dex when FWHM is used in the virial product (top), and
$0.204\pm0.011$ dex for the \sigmaline-based virial product (bottom).  
The average offset of all measurements based on mean
spectra is $0.147\pm0.066$ ($0.201\pm0.075$) dex when
\FWHM\ (\sigmaline) is used in the virial product.  To avoid potential 
biases from the narrowest-line object, we recalculate the average
offset after removing NGC\,4253.  The average offset of all SE
measurements is now $0.093\pm0.006$ dex and $0.135\pm0.007$ dex for
\FWHM\ and \sigmaline, respectively. When the measurements from mean
spectra are used in comparison, the average offsets are
$0.092\pm0.040$ dex and $0.136\pm0.043$ dex for \FWHM\ and
\sigmaline, respectively.  
Thus, the SE BH 
masses of broad-line AGNs with virial products in the range 
$\sim$10$^{5-7}$ M$_{\odot}$ can be overestimated 
by $\sim$25--35\% if the same recipe used for rms spectra is adopted.

These results are similar to the findings by Collin et al. (2006),
who investigated the systematic offset of virial product estimates 
between rms and mean spectra using a different sample of reverberation mapping measurements.
Although it is not straightforward to directly compare 
their results with ours since 
the methods of generating rms spectra and measuring line widths are
substantially different, the similar systematic offset between SE and 
reverberation masses clearly demonstrates the importance of calibrating SE masses.

%%%%%%%%%%%%%%%%%%%%%%
% line profile fit
%%%%%%%%%%%%%%%%%%%%%%
\begin{figure}
\centering
    \includegraphics[width=0.4\textwidth]{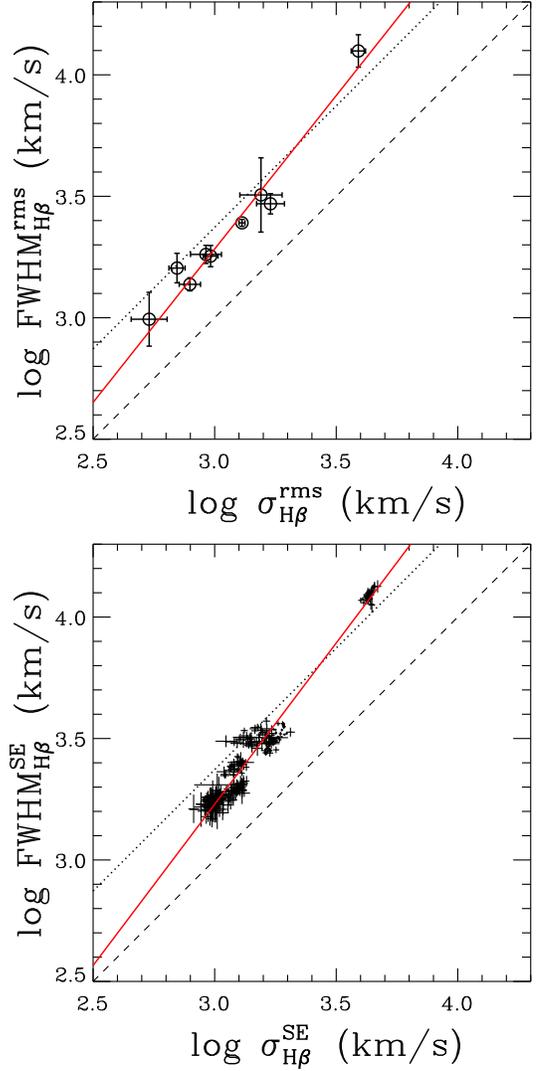}
    \caption{Comparisons of \FWHM\ and \sigmaline, using measurements from
    rms spectra (top) and SE spectra (bottom). The dashed line indicates 
    an identity relationship while the dotted line
    represents the Gaussian profile, i.e., \FWHM/\sigmaline\ $= 2.35$.
    The red solid line shows the best-fit relation.
    \label{fig:FWHM_sigma}}
\end{figure}

\subsubsection{Comparing \FWHM~ and \sigmaline}

We compare \FWHM\ and \sigmaline\ in Figure 15.  As previously noticed
in other studies (e.g., Collin et al. 2006; McGill et al.\ 2008), the
shape of the \Hb\ line is different from a Gaussian profile, for which
\FWHM/\sigmaline\ is expected to be 2.35. As shown in Figure 16, narrower
lines tend to have stronger wings leading to a lower \FWHM/\sigmaline\
ratio, while broader lines are more core dominated with a higher
\FWHM/\sigmaline\ ratio. These results are consistent with those of previous
studies, although our sample is composed of objects with narrower
lines than previously studied.

The best-fit correlation based on the rms spectra is
%log SIGMA(rms) = 0.40073081(+/-0.39496845) + 0.79173643(+/-0.11713576) log FWHM(rms)  (I.S. 0.058425434+/-0.042806968)
\begin{equation}
  \log {\rm \sigma_{\rm H\beta}}({\rm rms})  =  0.401 (\pm0.395)  + 0.792 (\pm0.117)~\log{\rm FWHM_{\rm H\beta}}({\rm rms}).
\label{eqn:convert_velrms}
\end{equation}
In the case of measurements from the SE spectra, the best fit is expressed as
% log SIGMA(SE) = 0.56736522(+/-0.026983573) + 0.75328588(+/-0.007780917)4 log FWHM(SE)  (I.S. 0.035627319+/-0.0015684047)
\begin{equation}
  \log {\rm \sigma_{\rm H\beta}}({\rm SE})  =  0.567 (\pm0.027) + 0.753 (\pm0.008)~\log{\rm FWHM_{\rm H\beta}}({\rm SE}).
\label{eqn:convert_velSE}
\end{equation}
We note that these results are somewhat limited by the small dynamic
range of our sample and the lack of objects with FWHM $>3000$~\kms. 
Further analysis with broader line objects is required.
However, in the case of broader line objects in the literature, 
we do not have
consistently measured line widths from rms spectra as
described in \S 3.3. Nevertheless, we will use this fit to convert
\FWHM\ to \sigmaline\ in \S 4.3.4.

\subsubsection{Line-Width Dependent Mass Estimators}

In order to avoid potential systematic biases in SE spectra, we derive
line-width dependent mass estimators, using the best-fit relations
derived above (see Fig.~\ref{fig:offset_vel}).  As a reference, we use
the mass estimator normalized for the virial product from rms spectra
(reverberation results) using the virial factor $\log~f=0.72$
determined from the
\msigma\ relation of reverberation mapped AGNs (Woo et al. 2010):

\begin{equation}
M_{\rm BH} = 10^{7.602}{\rm M}_{\odot} \left(
{\sigma_{{\rm H}\beta}({\rm rms}) \over 1000~ {\rm km~s}^{-1} } \right)^{2} \left( {\lambda
L_{5100,n} \over 10^{44}~ {\rm erg~s}^{-1}} \right)^{0.518}~.
\label{eq:MBH_sigmaRMS}
\end{equation}
If we replace \sigmaline\ from rms spectra with
\sigmaline\ measured from SE spectra using Eq. \ref{eqn:convert_velsigma}, 
the mass estimator changes to
\begin{equation}
M_{\rm BH} = 10^{7.370}{\rm M}_{\odot} \left(
{\sigma_{{\rm H}\beta}({\rm SE}) \over 1000~ {\rm km~s}^{-1} } \right)^{2.212} \left( {\lambda
L_{5100,n} \over 10^{44}~ {\rm erg~s}^{-1}} \right)^{0.518}~.
\label{eq:MBH_sigmaSE}
\end{equation}
%for AGN with \sigmaline~$<$12,000 \kms. 
As in the case of Eq. \ref{eqn:convert_velsigma}, we recommend readers use Eq. \ref{eq:MBH_sigmaSE}
for AGNs with \sigmaline < 2,000 \kms.

In the case of \FWHM, the virial factor has not been determined by Woo et
al. (2010), but we can use the relations found above to derive a
consistent expression. If we replace \sigmaline\ with \FWHM\ using
Eq. \ref{eqn:convert_velrms}, then Eq. \ref{eq:MBH_sigmaRMS} becomes
\begin{equation}
M_{\rm BH} = 10^{7.156}{\rm M}_{\odot} \left(
{{\rm FWHM}_{{\rm H}\beta}({\rm rms}) \over 1000~ {\rm km~s}^{-1} } \right)^{1.584} \left( {\lambda
L_{5100,n} \over 10^{44}~ {\rm erg~s}^{-1}} \right)^{0.518}~.
\label{eq:MBH_fwhmRMS}
\end{equation}

In order to use \FWHM\ measured from SE spectra, \FWHM\ from rms
spectra in Eq. \ref{eq:MBH_fwhmRMS} can be replaced by \FWHM\ from SE
spectra, using Eq. \ref{eqn:convert_velFWHM}.  Then, the mass
estimator becomes
\begin{equation}
M_{\rm BH} = 10^{6.966}{\rm M}_{\odot} \left(
{{\rm FWHM}_{{\rm H}\beta}({\rm SE}) \over 1000~ {\rm km~s}^{-1} } \right)^{1.734} \left( {\lambda
L_{5100,n} \over 10^{44}~ {\rm erg~s}^{-1}} \right)^{0.518}~.
\label{eq:MBH_fwhmSE}
\end{equation}

Alternatively, we can use Eq. \ref{eq:MBH_sigmaSE} and replace \sigmaline\ with 
\FWHM\ measured from SE spectra using Eq. \ref{eqn:convert_velSE}. Then Eq. \ref{eq:MBH_sigmaSE} becomes
\begin{equation}
M_{\rm BH} = 10^{6.985}{\rm M}_{\odot} \left(
{{\rm FWHM}_{{\rm H}\beta}(SE) \over 1000~ {\rm km~s}^{-1} } \right)^{1.666} \left( {\lambda
L_{5100,n} \over 10^{44}~ {\rm erg~s}^{-1}} \right)^{0.518}~,
\label{eq:MBH_fwhmSE_2}
\end{equation}
which is almost identical to Eq. \ref{eq:MBH_fwhmSE}. 
For a consistency check, we compared the SE masses estimated from 
Eq. \ref{eq:MBH_fwhmSE} with those from Eq. \ref{eq:MBH_fwhmSE_2}. 
They are consistent within $\sim$1\%, indicating that Eq. 15 and 16 are
essentially equivalent. 
As in the case of Eq. \ref{eqn:convert_velFWHM}, for AGNs with \FWHM < 3,000 \kms~
we recommend readers use Eq. \ref{eq:MBH_fwhmSE_2} instead of
Eq. \ref{eq:MBH_fwhmSE}, since the SE masses derived from Eq. 16 
are slightly more consistent with the masses determined from Eqs. 12 and 14.

The BH masses derived from the new mass estimators
are consistent with each other within a $\sim$2\% offset,
indicating that the systematic difference in the line widths between SE 
and rms spectra is well calibrated.
In contrast, the $\sim$0.2 dex scatter between various mass estimators 
reflects a lower limit to the uncertainties of our line-width dependent calibrations.
In a sense, these new estimators can be thought of as introducing a
line-width dependent virial factor to correct for the systematic
difference of the geometry and kinematics of the gas contributing to
the SE line profile and that contributing to the rms
spectra. Regardless of the physical interpretation, these new recipes
ensure that mass estimates from SE spectra and rms spectra can be
properly compared.

%%%%%%%%%%%%%%%%%%%%%%%%%%%%%%%%%%%%%%%%%%%%%%%%%%%%%%%%%%%%%%%%%%%%%%%%%%%
% Discussion and Conclusion
%%%%%%%%%%%%%%%%%%%%%%%%%%%%%%%%%%%%%%%%%%%%%%%%%%%%%%%%%%%%%%%%%%%%%%%%%%%
\section{DISCUSSION AND CONCLUSIONS}  \label{section:diss}
\subsection{Random Uncertainty}

We investigated the precision and accuracy of BH mass estimates based
on SE spectra, using the homogeneous and high-quality spectroscopic
monitoring of 9 local Seyfert 1 galaxies obtained as part of the LAMP
project.
% now the random errors are changed from 0.1 dex (using R-L relation)
% to 0.05 dex (using decomposition Ln)
We find that the uncertainty of SE mass estimates due to the AGN
variability is $\sim 0.05$ dex ($\sim$12\%). Our result is slightly
less than that of \citet{denney09}, who reported $\sim0.1$ dex random
error due to the variability based on the investigation of Seyfert 1
galaxy NGC\,5548 using data covering $\sim$10 years.  For higher
luminosity AGNs, the uncertainty due to variability can be smaller
since the amplitude of variability inversely correlates with the
luminosity (e.g., Cristiani et al. 1997). For example, by comparing SE
spectra with mean spectra averaged over $\sim$10 multi-epoch data of 8
moderate-luminosity AGN, Woo et al. (2007) reported that intrinsic
FWHM variation of the \Hb\ line is $\sim$7\%, resulting in $\sim$15\%
random error in mass estimates.

In addition to the uncertainty related to variability, the total
random uncertainty of SE mass estimators includes the uncertainty in
the virial factor, and the scatter of the size-luminosity relation.
The scatter of the AGN \msigma\ relation (Woo et al. 2010) provides an
upper limit to the random object-to-object scatter in the virial
factor of 0.43 dex. By adding 0.1 dex due to variability and 0.13 dex
scatter from the size-luminosity relation in quadrature (assuming they
are uncorrelated), the upper limit of the overall uncertainty of SE
mass estimates is found to be 0.46 dex. This is consistent with the
uncertainty of 0.4--0.5 dex estimated by Vestergaard \& Peterson
(2006). If we assume more realistically that 0.3 dex of the scatter in
the \msigma\ relation measured by Woo et al. (2010) is intrinsic
scatter (e.g., G\"ultekin et al. 2010) and not due to uncertainties in
the virial coefficient, then the uncertainty of the virial factor
becomes 0.31 dex, resulting in an overall uncertainty of $\sim$0.35
dex in SE mass estimates. More direct measurements of the virial
coefficient (e.g., Davies et al. 2006; Onken et al. 2007; Hicks \& Malkan 2008; Brewer et al. 2011) are needed to break this
degeneracy.

Note that measurement errors in the line width and continuum
luminosity are negligible in our study owing to the high quality of
the data. However, often such high-quality data are not available, and
measurement errors of the line width in particular can be a
significant contribution to the total error budget.  For example, Woo
et al. (2007) estimated the propagated uncertainty in the SE mass
estimates due to the FWHM measurement errors as 0.11 dex (30\%) based
on spectra with a S/N of 10--15.  Therefore, the estimated overall
uncertainty of $\sim$0.35 dex should be taken as a lower limit for
typical SE mass estimates based on optical spectra such as those from
the Sloan Digital Sky Survey.

%%%%%%%%%%%%%%%%%%%%%%
\begin{figure*}
\centering
    \includegraphics[width=0.4\textwidth]{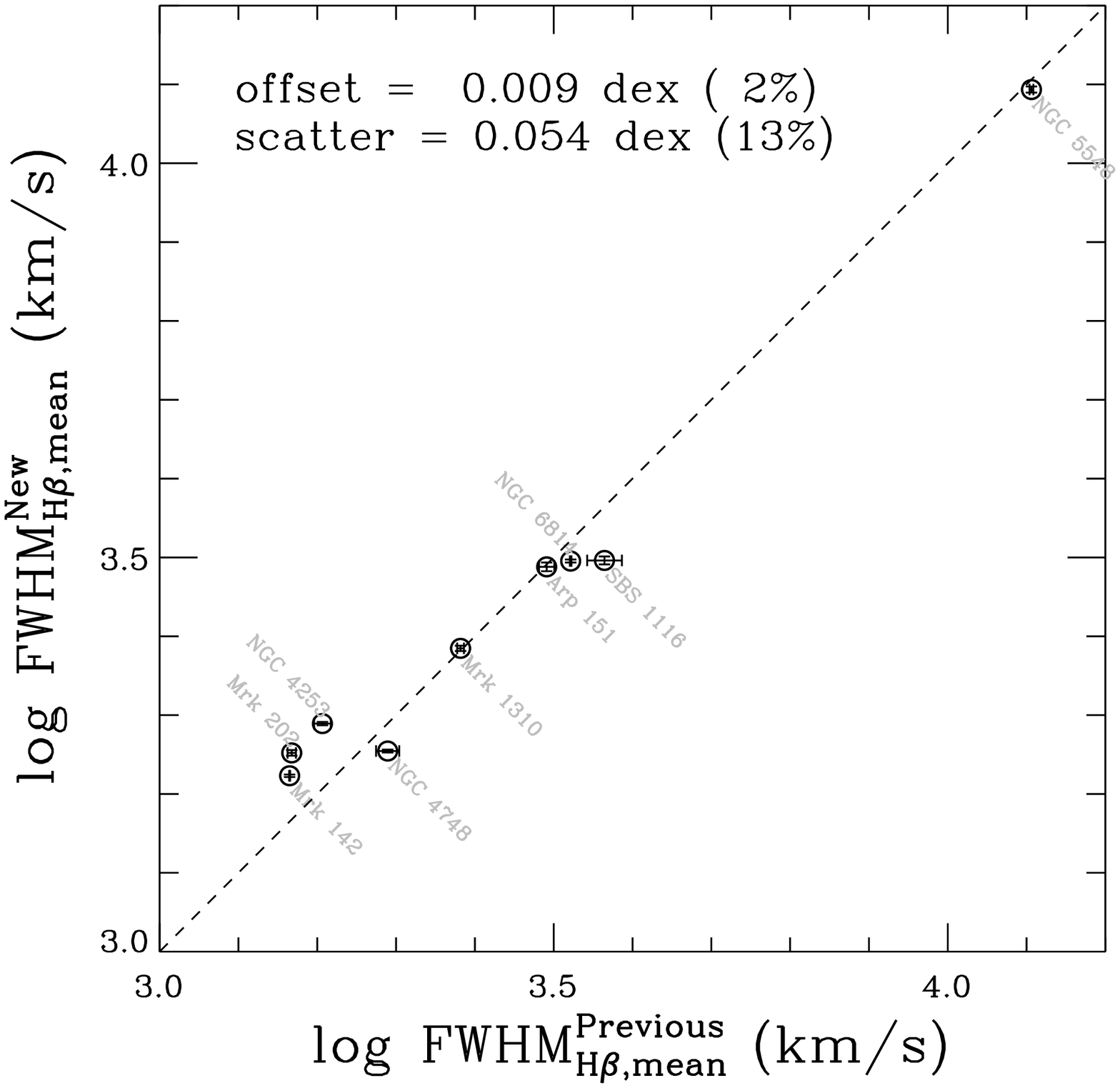}  \hspace{2em}
    \includegraphics[width=0.4\textwidth]{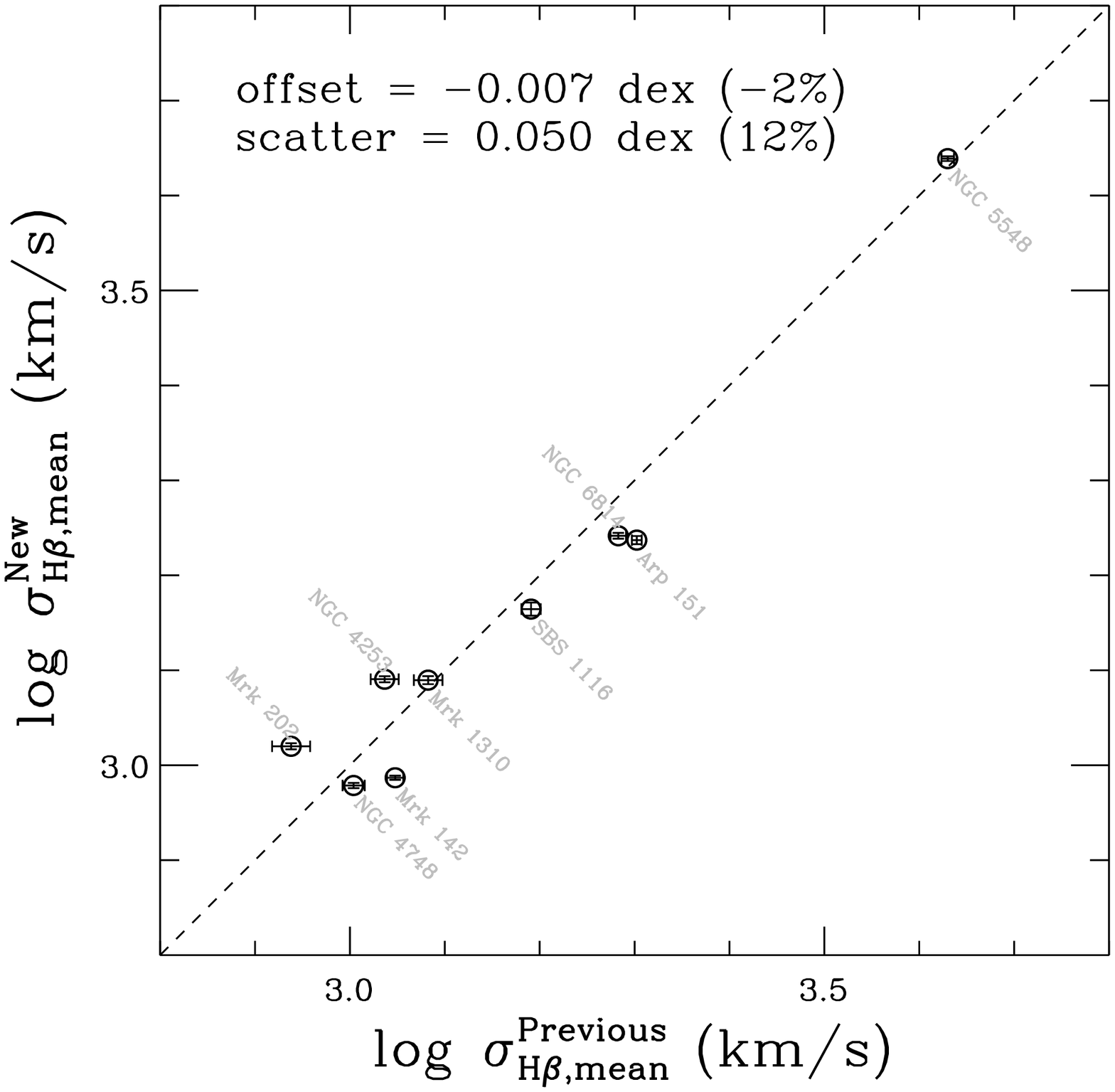}  \\
    \includegraphics[width=0.4\textwidth]{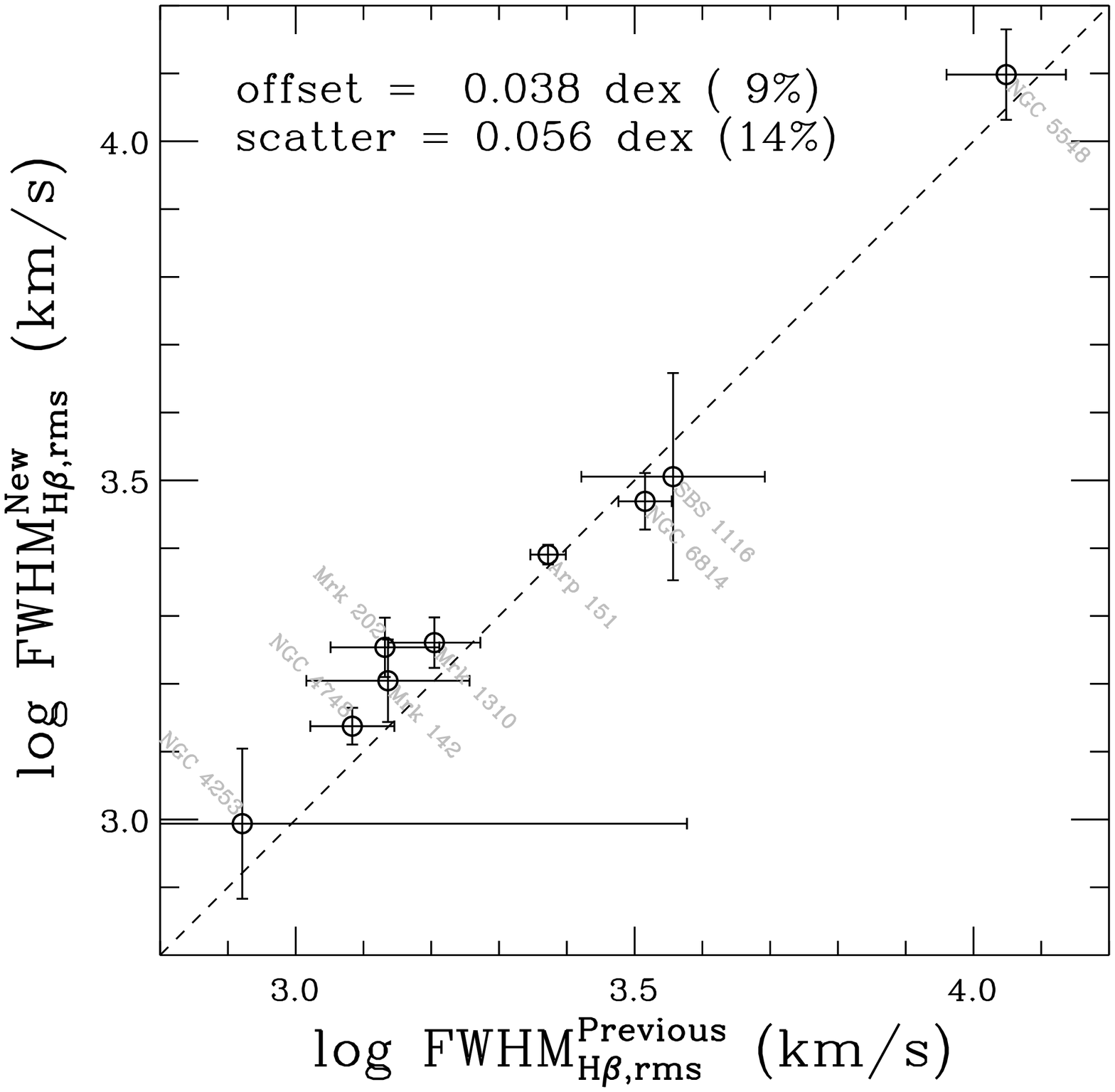}  \hspace{2em}
    \includegraphics[width=0.4\textwidth]{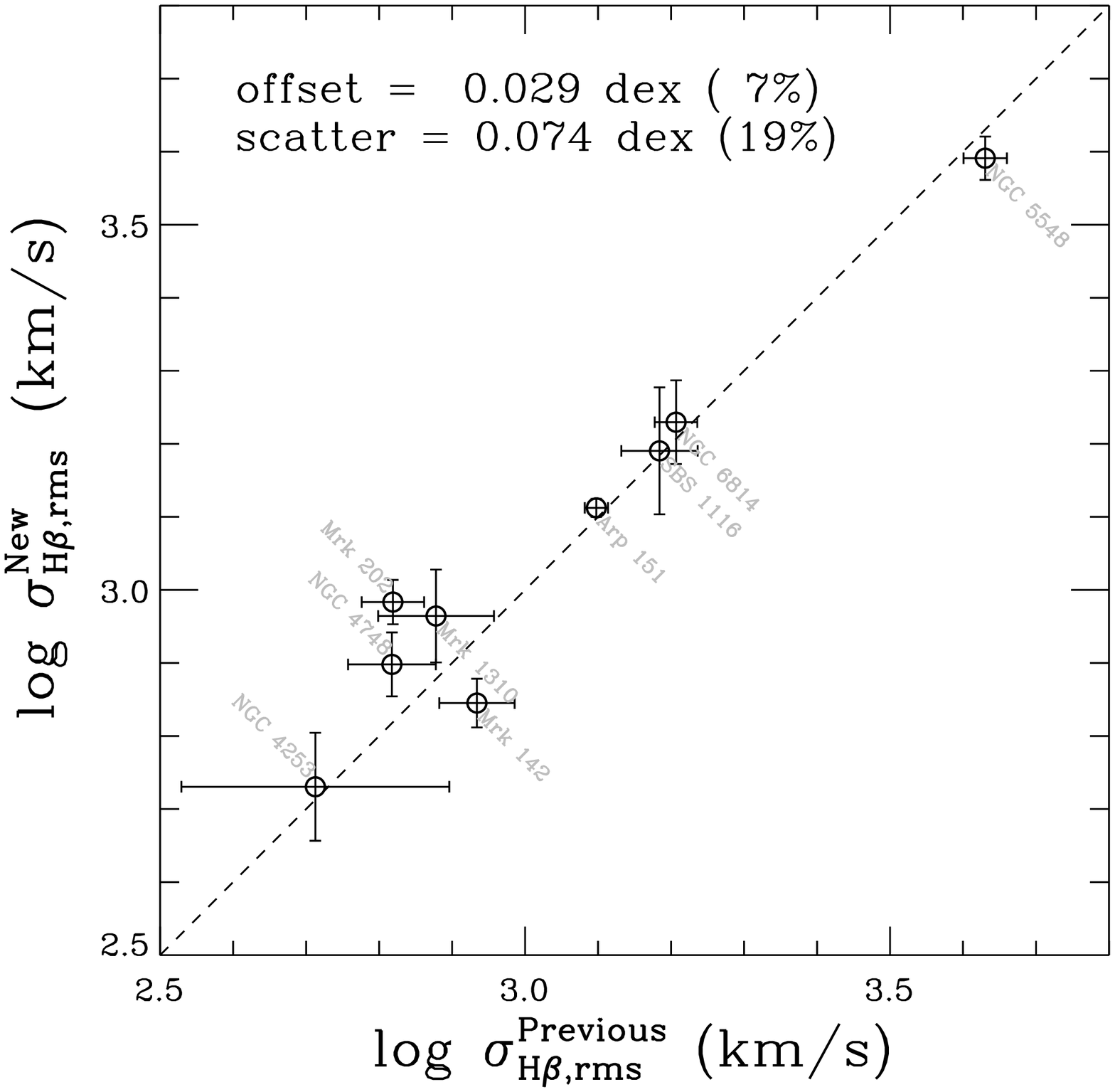}
    \caption{Comparison of \FWHM~(left) and \sigmaline~measurements (right)
    based on the new methods (this work) and the previous traditional methods (Bentz et al. 2009c),
    using the mean (top) and rms spectra (bottom). The average difference
    of the \Hb~line widths between two methods is relatively small. 
    However, the large scatter (12--19\%) indicates
    that the contribution from individual effects, e.g., subtraction of starlight,
    \ion{Fe}{2}, or \ion{He}{2} varies on an object-by-object basis.} 
\end{figure*}

\subsection{Difference in Line Profile between SE and RMS spectra}

We confirmed that the \Hb\ line width measured from a mean or SE
spectrum is systematically larger than that measured from an rms
spectrum.  The systematic difference corresponds to an average
difference in virial product of $\sim$0.15--0.20 dex.  
However, the
average difference is dominated by the narrowest line objects in the
sample, with a decreasing trend toward broader line objects as shown in Fig. \ref{fig:offset_mass}
(cf., Collin et al. 2006).
These results indicate that for narrow-line AGNs (\FWHM\ $< 3000$~\kms), 
BH masses based on SE spectra can be
overestimated by $\sim$25--35\% if standard recipes are used. 
In order to correct for the systematic difference of the line profile, we
derive new empirically calibrated line-width dependent SE mass
estimators.

It is important to notice that line-width measurements from
rms spectra can also be systematically biased by residuals of
narrow-line components, \ion{Fe}{2}, and host-galaxy starlight. Fluctuations
of these components can generate signatures in the rms spectra,
resulting in a biased line profile and an improper continuum fit. We
have demonstrated several new strategies to mitigate these
effects. First, we adopt two robust methods to derive rms spectra
--- using S/N weights and adopting a maximum likelihood approach ---
that minimize the contamination of low-S/N spectra when there is
a large range in S/N. Second, we
subtract \ion{Fe}{2} and host-galaxy starlight based on spectral
decomposition analysis of each individual-epoch spectrum before making
the rms and mean spectra. 
These new methods substantially improve the quality of rms spectra
in measuring the width of the \Hb~line 
for AGNs with strong starlight, \ion{Fe}{2}, or blended \ion{He}{2}.
Especially in these cases we recommend the new methods as an useful alternative to 
the traditional simple methods in future reverberation mapping studies 
(and possibly to revisit previous studies). 

The new methods introduced here differ from previous ones. The 
overall goal is to account for systematic uncertainties and correct 
for biases stemming from known effects such as stellar line contamination. 
However, it is possible that they might introduce biases due to unknown 
systematics, especially when directly compared to other measurements 
obtained with previous techniques. An absolute comparison would require 
a third way to measure the same quantities (e.g., BH mass); 
however, we can estimate any differential bias by comparing our measurements 
to those given by \citet{bentz09c} using traditional methods.
On average, \sigmaline\ measured with our new methods increases 
by $7\pm6$\% and \FWHM\ increases by $9\pm4$\% (see Fig. 17),
suggesting that the systematic uncertainties due to the new methods
%including removal of host galaxy starlight and \ion{Fe}{2},
is lower than 10\%, although we assume the difference is entirely caused by the
systematic uncertainty of the new scheme.

The average small difference between our new measurements and those 
of \citet{bentz09c} is due to the fact that various effects, e.g.,
S/N weighting, prior subtraction of host galaxy starlight 
and blended emission lines, line profile fitting, and removal of narrow \Hb, 
are mixed together and canceled out for individual objects. 
For example, if we separate the effect of host galaxy starlight,
the difference of the line dispersion measurements
with/without removal of host galaxy starlight is much larger than 
the average difference between \citet{bentz09c} and ours (see \S 3.3.2).               
The 10-20\% rms scatter between previous \citep{bentz09c} and new measurements
indicates that the contribution of each effect varies on an object-by-object basis
(see Fig. 17). 
The new methods are useful for reducing the uncertianites of individual
BH masses due to those various effects, and better constraining the intrinsic
scatter of the \msigma~relation (e.g., Woo et al. 2010).
When other systematic uncertainites, i.e., the virial coefficient,
can be constrained and reduced in the future, 
the new methods will become more important for BH mass determination.

From the point of view of interpretation, the systematically narrower
line width in the rms spectra can be explained by the photoionization
calculations of \citet{KG04}, which predict that high-velocity gas in
the inner BLR has lower responsivity, leading to lower variability of
the line wings and therefore a narrower profile in the rms
spectrum. However, it is not clear why the effect should be stronger
for the narrower line AGNs. Further investigations are required to
reveal the nature of this systematic trend.  We conclude by noting
that our study is limited to relatively low luminosity and narrow line
width (and hence small BH mass) AGNs.  In future work we plan to
expand our study to cover a larger dynamic range in luminosity and
line width.

\subsection{Implications for the Evolution of BH Host-Galaxy Scaling Relations}

Virtually all observational studies of the evolution of the BH
host-galaxy scaling relations over cosmic time are based on SE mass
estimates.  Evolutionary trends are generally established by comparing
measurement of distant samples (based on SE BH mass estimates) with
the local scaling relations. The latter can be determined either based
on SE BH mass estimates, or on reverberation BH mass estimates, or on
spatially resolved kinematics. In the case of comparison between SE
mass estimates at high redshift and mass estimates of local objects
based on different methods, it is essential that the mass estimates be
properly calibrated. For example, the positive bias of \Hb\ line width
of SE spectra compared to the rms spectra could lead to overestimation
of BH masses for distant samples, if compared with local samples based
on rms spectra.

However, as shown in Figures 13 and 14, this bias is only significant
for the narrower line objects with BH mass $<10^7$
M$_\odot$). Typically, high-redshift studies such as those by
\citet{bennert10} and \citet{Merloni10} focus on higher mass BHs,
where the bias is believed to be negligible. We note that
one could completely eliminate this bias by comparing distant and
local BH mass estimates based entirely on self-consistent SE BH mass
estimates (\citealt{woo08};\citealt{bennert11a,bennert11b}).  
Even then, of course, one must keep in mind the differential nature of the
measurement. For example, the slope inferred for the local scaling
relations based on SE spectra (e.g., \citealt{GH06}) may be biased
with respect to the true slope if the mass estimator is biased at low
masses, and yet one may still infer the correct evolution even for low
masses if the bias does not change with redshift.

%%%%%%%%%%%%%%%%%%%%%%%%%%%%%%%%%%%%%%%%%%%%%%%%%%%%%%%%%%%%%%%%%%%%%%%%%%%
% Acknowledgments
%%%%%%%%%%%%%%%%%%%%%%%%%%%%%%%%%%%%%%%%%%%%%%%%%%%%%%%%%%%%%%%%%%%%%%%%%%%
\acknowledgments

This work has been supported by the Basic Science Research Program through 
the National Research Foundation of Korea funded by the Ministry of Education, 
Science and Technology (2010-0021558). We acknowledge NSF grants 
AST-0548198 (UC Irvine), AST-0908886 and AST-1108665 (UC Berkeley), 
AST-0642621 (UC Santa Barbara), and AST-0507450 (UC Riverside).
We are grateful to the staff at Lick Observatory for their assistance
with the observations. We thank the anonymous referee for constructive 
comments and suggestions.
%D. Park would like to thank Hyung Mok Lee, Matthew W. Auger, and Jong Chul Lee for helpful discussions.

\   

%%%%%%%%%%%%%%%%%%%%%%%%%%%%%%%%%%%%%%%%%%%%%%%%%%%%%%%%%%%%%%%%%%
%                       REFERENCES
%%%%%%%%%%%%%%%%%%%%%%%%%%%%%%%%%%%%%%%%%%%%%%%%%%%%%%%%%%%%%%%%%%


\begin{thebibliography}{dummy}
\expandafter\ifx\csname natexlab\endcsname\relax\def\natexlab#1{#1}\fi
\bibitem[Barth et al.(2005)]{barth05}
Barth, A.~J., Greene, J.~E., \& Ho, L.~C., 2005, \apj, 619, L151
\bibitem[Barth et al.(2011)]{barth11}
Barth, A. J., et al. 2011, \apj, 732, 121
\bibitem[Bennert et al.(2010)]{bennert10}
Bennert, V. N., et al. 2010, \apj, 708, 1507
\bibitem[Bennert et al.(2011{\natexlab{a}})]{bennert11a}
Bennert, V. N., et al. 2011{\natexlab{a}}, \apj, 726, 59
\bibitem[Bennert et al.(2011{\natexlab{b}})]{bennert11b}
Bennert, V. N., et al. 2011{\natexlab{b}}, arXiv:1102.1975
\bibitem[{{Bentz} {et~al.}(2006){Bentz}, {Peterson}, {Pogge},
  {Vestergaard}, \& {Onken}}]{bentz06}
{Bentz}, M.~C., {Peterson}, B.~M., {Pogge}, R.~W., {Vestergaard}, M., \&
  {Onken}, C.~A. 2006, \apj, 644, 133
\bibitem[Bentz et al.(2009{\natexlab{a}})]{bentz09a}
{Bentz}, M.~C., {Peterson}, B.~M., {Pogge}, R.~W., \& {Vestergaard}, M. 2009{\natexlab{a}}, \apj, 694, L166
\bibitem[Bentz et al.(2009{\natexlab{b}})]{bentz09b}
Bentz, M. C., et al. 2009{\natexlab{b}}, \apj, 697, 160
\bibitem[Bentz et al.(2009{\natexlab{c}})]{bentz09c}
Bentz, M. C., et al. 2009{\natexlab{c}}, \apj, 705, 199
%\bibitem[Bentz et al.(2010)]{bentz10}
%Bentz, M. C., et al. 2010, \apj, 716, 993
\bibitem[Brewer et al.(2011)]{Brewer11}
Brewer, B. J., et al. 2011, \apj, 733, L33
\bibitem[{{Blandford} \& {McKee}(1982)}]{BM82}
{Blandford}, R.~D., \& {McKee}, C.~F. 1982, \apj, 225, 419
\bibitem[{{Boroson} \& {Green}(1992)}]{Boroson92}
{Boroson}, T.~A., \& {Green}, R.~F. 1992, \apjs, 80, 109
\bibitem[{{Bruzual} \& {Charlot}(2003)}]{Bruzual03}
{Bruzual}, G., \& {Charlot}, S. 2003, \mnras, 344, 1000
\bibitem[{{Christiani} {et~al.}(1997)}]{christiani97}
{Cristiani}, S., {Trentini}, S., {La Franca}, F., \& {Andreani}, P.
1997, \aap, 321, 123
\bibitem[{{Collin} {et~al.}(2006)}]{collin06}
{Collin}, S., {Kawaguchi}, T., {Peterson}, B.~M., \& {Vestergaard},
M. 2006, \aap, 456, 75

\bibitem[Davies et al.(2006)]{davies+06} Davies, R.~I., et al.\ 
2006, \apj, 646, 754

\bibitem[{{Davis} {et~al.}(2007)}]{davis07}
{Davis}, S.~W., {Woo}, J.-H., \& {Blaes}, O.~M., 2007, \apj, 668, 682 
%\bibitem[{{Decarli} {et~al.}(2008){Decarli}, {Labita}, {Treves}, \&
%  {Falomo}}]{decarli08}
%{Decarli}, R., {Labita}, M., {Treves}, A., \& {Falomo}, R. 2008,
%\mnras, 387, 1237
\bibitem[{{Decarli} {et~al.}(2010){Decarli}, {Labita}, {Treves}, \&
  {Falomo}}]{decarli08}
{Decarli}, R., {Falomo}, R., {Treves}, A., {Labita}, M., {Kotilainen}, J. K., \& {Scarpa}, R., 
2010, \mnras, 402, 2453
\bibitem[{{Denney} {et~al.}(2009)}]{denney09}
{Denney}, K.~D., {et~al.} 2009, \apj, 692, 246
\bibitem[{{Denney} {et~al.}(2010)}]{denney10}
{Denney}, K.~D., {et~al.} 2010, \apj, 721, 715
\bibitem[{{Dietrich} {et~al.}(2005)}]{diet05}
{Dietrich}, M., {Crenshaw}, D.~M., \& {Kraemer}, S.~B., 2005, \apj, 623, 700
\bibitem[{{Ferrarese} \& {Ford}(2005)}]{FF05}
{Ferrarese}, L., \& {Ford}, H. 2005, \ssr, 116, 523
\bibitem[{{Ferrarese} \& {Merritt}(2000)}]{FM00}
{Ferrarese}, L., \& {Merritt}, D. 2000, \apj, 539, L9
\bibitem[{{Fine} {et~al.}(2008)}]{F+00}
{Fine}, S., {et~al.} 2008, \mnras, 390, 1413
\bibitem[{{Gebhardt} {et~al.}(2000)}]{G+00}
{Gebhardt}, K., {et~al.} 2000, \apj, 539, L9
\bibitem[{{Greene} \& {Ho}(2006)}]{GH06}
{Greene}, J.~E., \& {Ho}, L.~C. 2006, \apj, 641, 21
\bibitem[{{G\"ultekin} {et~al.}(2009)}]{Gul+09}
{Gultekin}, K., {et~al.} 2009, \apj, 698, 198
\bibitem[Hicks 
\& Malkan(2008)]{2008ApJS..174...31H} Hicks, E.~K.~S., \& Malkan, M.~A.\ 2008, \apjs, 174, 31 

\bibitem[{{Hopkins} {et~al.}(2006)}]{hopkins+09}
{Hopkins}, P.~F., {et~al.} 2006, \apjs, 163, 1
\bibitem[{{Kaspi} {et~al.}(2005){Kaspi}, {Maoz}, {Netzer}, {Peterson},
  {Vestergaard}, \& {Jannuzi}}]{kaspi05}
{Kaspi}, S., {Maoz}, D., {Netzer}, H., {Peterson}, B.~M.,
{Vestergaard}, M., \& {Jannuzi}, B.~T. 2005, \apj, 629, 61
\bibitem[{{Kaspi} {et~al.}(2000){Kaspi}, {Smith}, {Netzer}, {Maoz}, {Jannuzi},
  \& {Giveon}}]{kaspi00}
{Kaspi}, S., {Smith}, P.~S., {Netzer}, H., {Maoz}, D., {Jannuzi}, B.~T., \&
  {Giveon}, U. 2000, \apj, 533, 631
\bibitem[{{Kauffmann} \& {Haehnelt}(2000)}]{KH00}
{Kauffmann}, G., \& {Haehnelt}, M. 2000, \mnras, 311, 576
\bibitem[{{Kelly}(2007)}]{kelly03}
{Kelly}, B. C., 2007, \apj, 665, 1489
\bibitem[{{Kollatschny}(2003)}]{kolla03}
{Kollatschny}, W., 2003, \aap, 407, 461
\bibitem[{{Kollmeier}(2006)}]{kollm06}
{Kollmeier}, J., A., 2006, \apj, 648, 128
\bibitem[{{Kormendy} \& {Richstone}(1995)}]{KR95}
{Kormendy}, J., \& {Richstone}, D. 1995, \araa, 33, 581
\bibitem[Kormendy \& Gebhardt(2001)]{KG01} Kormendy, J., \& Gebhardt, K.\ 
2001, in 20th Texas Symposium on Relativistic Astrophysics, ed. J. C.
Wheeler \& H. Martel (New York: AIP, vol. 586), 363 
\bibitem[{{Korista} \& {Goad}(2004)}]{KG04}
{Korista}, K.~T., \& {Goad}, M.~R., 2004, \apj, 606, 749
\bibitem[{{Lauer} {et~al.}(2007)}]{lauer07}
{Lauerrconi}, T.,~R., et al.\ 2007, \apj, 670, 249
\bibitem[{{Marconi} \& {Hunt}(2003)}]{MH03}
{Marconi}, A., \& {Hunt}, L. K. 2003, \apj, 589, L21
\bibitem[{{Marconi} {et~al.}(2008)}]{marconi08}
{Marconi}, A., et al. 2008, \apj, 678, 693
\bibitem[{{Magorrian} {et~al.}(1998)}]{M+98}
Magorrian, J. et al.\ 1998, \aj, 115, 2285
\bibitem[{Markwardt(2009)}]{Markwardt09}
Markwardt, C.~B. 2009, in Astronomical Data
Analysis Software and Systems XVIII, ed. D. A. Bohlender, D. Durand, \&
P. Dowler (San Francisco: ASP), 251
\bibitem[{{McGill} {et~al.}(2008)}]{mcgill08}
{McGill}, K.~L., {Woo}, J.-H., {Treu}, T., \& {Malkan}, M.~A. 2008,
\apj, 673, 703
\bibitem[{{McLure} \& {Dunlop}(2004)}]{mclure04}
{McLure}, R.~J., \& {Dunlop}, J.~S. 2004, \mnras, 352, 1390
\bibitem[{{Merloni} {et~al.}(2010)}]{Merloni10}
{Merloni}, A. et~al. 2010, \apj, 708, 137
\bibitem[{{Onken} {et~al.}(2004){Onken}, {Ferrarese}, {Merritt}, {Peterson},
  {Pogge}, {Vestergaard}, \& {Wandel}}]{onken04}
{Onken}, C.~A., {Ferrarese}, L., {Merritt}, D., {Peterson}, B.~M.,
{Pogge},
  R.~W., {Vestergaard}, M., \& {Wandel}, A. 2004, \apj, 615, 645
\bibitem[{{Netzer} \& {Marziani}(2010)}]{NM10}
{Netzer}, H., \& {Marziani}, P. 2010, \apj, 724, 318

\bibitem[Onken et al.(2007)]{2007ApJ...670..105O} Onken, C.~A., et al.\ 
2007, \apj, 670, 105

\bibitem[{{Pancoast} {et~al.}(2011)}]{Pan11}
{Pancoast}, A., {Brewer}, B.~J., \& {Treu}, T. 2011, \apj, 730, 139
\bibitem[{Peng} {et~al.}(2006)]{P06} Peng, C.~Y., Impey, C.~D., 
Rix, H.-W., Kochanek, C.~S., Keeton, C.~R., Falco, E.~E., Leh{\'a}r, J., 
\& McLeod, B.~A.\ 2006, \apj, 649, 616 
\bibitem[{{Peterson}(1993)}]{P93}
{Peterson}, B.~M., 1993, \pasp, 105, 247
\bibitem[{{Peterson} \& {Wandel}(1999)}]{PW99}
{Peterson}, B.~M., \& {Wandel}, A., 1999, \apj, 521, L95
\bibitem[{{Peterson} \& {Wandel}(2000)}]{PW00}
{Peterson}, B.~M., \& {Wandel}, A., 2000, \apj, 540, L13
\bibitem[{{Peterson} {et~al.}(2004)}]{peterson04}
{Peterson}, B.~M., {et~al.} 2004, \apj, 613, 682
%\bibitem[{{Peterson}(2010)}]{P10}
%{Peterson}, B.~M., 2010, IAUS, 267, 151
%\bibitem[{{Rafiee} \& {Hall}(2010)}]{RH10}
%{Rafiee}, A., \& {Hall}, P.~B., 2010, arXiv:1011.1268
%\bibitem[{{Richstone} {et~al.}(1998)}]{R+98}
%Richstone, D. et al.\ 1998, \nat, 395, A14
\bibitem[{{Robertson} {et~al.}(2006)}]{robert06}
Robertson, B. et al. 2006, \apj, 641, 90
\bibitem[{{Sergeev} {et~al.}(1999)}]{serg99}
Sergeev, S. G., Pronik, V. I., Sergeeva, E. A., Malkov, Yu. F.
{Sergeev}, S.~G., {Pronik}, V.~I., {Sergeeva}, E.~A., \& {Malkov}, Yu.~F., 1999, \aj, 118, 2658
\bibitem[{{Shapovalova} {et~al.}(2004)}]{shapo04}
Shapovalova, A.~I., et al. 2004, \aa, 422, 925
\bibitem[{{Shen} {et~al.}(2008)}]{shen08}
{Shen}, Y., {Greene}, J.~E., {Strauss}, M.~A., {Richards}, G.~T., \&
  {Schneider}, D.~P. 2008, \apj, 680, 169
\bibitem[{{Shen} \& {Kelly}(2010)}]{SK10}
{Shen}, Y., \& {Kelly}, B.~C., 2010, \apj, 713, 41
\bibitem[{{Shen} {et~al.}(2011)}]{shen11}
Shen, Y., et al. 2011, \apjs, 194, 45
\bibitem[{{Shields} {et~al.}(1995)}]{shields10}
{Shields}, J.~W., {Ferland}, G.~J., \& {Peterson}, B.~M. 1995, \apj,
441, 507
\bibitem[{{Tremaine} {et~al.}(2002)}]{T+02}
{Tremaine}, S., {et~al.} 2002, \apj, 574, 740
\bibitem[{{Treu} {et~al.}(2007)}]{treu07}
{Treu}, T., {Woo}, J.-H., {Malkan}, M.~A., \& {Blandford}, R.~D. 2007, \apj,
667, 117
\bibitem[{{van Groningen} \& {Wanders}(1992)}]{vangroningen92}
{van Groningen}, E., \& {Wanders}, I. 1992, \pasp, 104, 700
\bibitem[{{van der Marel} \& {Franx}(1993)}]{vander93}
{van der Marel}, R.~P., \& {Franx}, M. 1993, \apj, 407, 525
\bibitem[{{Vanden Berk} {et~al.}(2001)}]{vanden01}
Vanden Berk, D.~E., et al.\ 2001, \aj, 122, 549
\bibitem[{{Vestergaard} \& {Peterson}(2006)}]{VP06}
{Vestergaard}, M., \& {Peterson}, B.~M. 2006, \apj, 641, 689
\bibitem[{{Wandel} {et~al.}(1999)}]{wandel99}
{Wandel}, A., {Peterson}, B.~M., \& {Malkan}, M.~A. 1999, \apj, 526, 579
\bibitem[{{Wilhite} {et~al.}(2007)}]{wilhite07}
{Wilhite}, B.~C., {Brunner}, R.~J., {Schneider}, D.~P., \& {Vanden
Berk}, D.~E. 2007, \apj, 669, 791
\bibitem[{{Woo} \& {Urry}(2002)}]{WU02}
{Woo}, J.-H., \& {Urry}, M., 2002, \apj, 579, 530
\bibitem[{{Woo} {et~al.}(2004){Woo}, {Urry}, {Lira},
  {van der Marel}, \& {Maza}}]{woo04}
{Woo}, J.-H., {Urry}, C.~M., {Lira}, P., {van der Marel}, R.~P.,  \&
{Maza}, J., 2004, \apj, 617, 903
\bibitem[{{Woo} {et~al.}(2006){Woo}, {Treu}, {Malkan}, \&
  {Blandford}}]{woo06}
{Woo}, J.-H., {Treu}, T., {Malkan}, M.~A., \& {Blandford}, R.~D.
2006, \apj, 645, 900
\bibitem[{{Woo} {et~al.}(2007)}]{woo07}
{Woo}, J.-H., {Treu}, T., {Malkan}, M.~A., {Ferry}, M.~A., \&
{Misch}, T. 2007, \apj, 661, 60
\bibitem[{{Woo} {et~al.}(2008)}]{woo08}
{Woo}, J.-H., {Treu}, T., {Malkan}, M.~A., \& {Blandford}, R.~D., 2008, \apj, 681, 925
\bibitem[{{Woo} {et~al.}(2010)}]{woo10}
{Woo}, J.-H.,  et al.\ 2010, \apj, 716, 269
\end{thebibliography}
\end{document}